\documentclass[prd,onecolumn,nofootinbib,showpacs]{revtex4-1}

\usepackage{bm}
\usepackage{amsmath}
\usepackage{natbib}
\usepackage{graphicx}
\usepackage{comment}

\def\ba{{\bm a}}
\def\bb{{\bm b}}

\def\bg{{\bm g}}

\def\bk{{\bm k}}

\def\bn{{\bm n}}
\def\br{{\bm r}}
\def\bN{{\bm N}}

\def\bv{{\bm v}}

\def\bx{{\bm x}}

\def\bR{{\bm R}}

\def\bX{{\bm X}}

\def\bbe{{\bm \beta}}

\def\aj{Astronomical Journal}
 
\def\planss{Planetary and Space Science}
\def\na{New A} 
\def\aap{A\&A}

\def\ba{{\bm a}}
\def\bb{{\bm b}}

\def\bk{{\bm k}}

\def\bn{{\bm n}}
\def\bv{{\bm v}}

\def\bx{{\bm x}}

\def\bz{{\bm z}}

\def\bN{{\bm N}}
\def\bR{{\bm R}}

\def\bX{{\bm X}}

\def\bbe{{\bm \beta}}
\def\lb{\label}
\def\bG{{\bm G}}

\def\bx{{\bm x}}  
\def\ba{{\bm a}}  
\usepackage[colorlinks=true]{hyperref}   

\begin{document}

\title{Light propagation in the field of a  moving axisymmetric body: theory and application to JUNO}

\author{A. Hees}
\email{A.Hees@ru.ac.za}
\affiliation{Department of Mathematics, Rhodes University, Grahamstown 6140, South Africa}

\author{S. Bertone\footnote{currently at Astronomical Institute, University of Bern, Siedlerstrasse 5, CH-3011 Bern, Switzerland}}
\email{stefano.bertone@obspm.fr}
\affiliation{Observatoire de Paris, SYRTE,CNRS/UMR 8630, UPMC \\ 61 Avenue de l'Observatoire, F-75014 Paris, France}

\author{C. Le Poncin-Lafitte}
\email{christophe.leponcin@obspm.fr}
\affiliation{Observatoire de Paris, SYRTE,CNRS/UMR 8630, UPMC \\ 61 Avenue de l'Observatoire, F-75014 Paris, France}

%\author{A. Hees$^{1}$, S. Bertone$^{2,}$\footnote{currently at Astronomical Institute, University of Bern, Siedlerstrasse 5, CH-3011 Bern, Switzerland} and C. Le Poncin-Lafitte$^2$}
%\affiliation{$^1$ Department of Mathematics, Rhodes University, Grahamstown 6140, South Africa}
%\affiliation{$^2$ Observatoire de Paris, SYRTE,CNRS/UMR 8630, UPMC \\ 61 avenue de l'Observatoire, F-75014 Paris, France}

\date{\today}

\begin{abstract}
Given the extreme accuracy of modern space science, a precise relativistic modeling of observations is required. We use the Time Transfer Functions formalism to study light propagation in the field of uniformly moving axisymmetric bodies, which extends the field of application of previous works. We first present a space-time metric adapted to describe the geometry of an ensemble of uniformly moving bodies. Then, we show that the expression of the Time Transfer Functions in the field of a uniformly moving body can be easily derived from its well-known expression in a stationary field by using a change of variables. We also give a general expression of the Time Transfer Function in the case of an ensemble of arbitrarily moving point masses. This result is given in the form of an integral easily computable numerically. We also provide the derivatives of the Time Transfer Function in this case, which are mandatory to compute Doppler and astrometric observables. We particularize our results in the case of moving axisymmetric bodies. Finally, we apply our results to study the different relativistic contributions to the range and Doppler tracking for the JUNO mission in the Jovian system.
\end{abstract}


\maketitle

%%%%%%%%%%%%%%%%%%%%%%%%%%%%%%%%%%%%%%%%%%%%%%%%%%%%%%%%%%%%%%%%%%%%%%%%%%
%%%%%%%%%%%%%%%%%%%%%%%%%%%%%%%%%%%%%%%%%%%%%%%%%%%%%%%%%%%%%%%%%%%%%%%%%%
\section{Introduction}
%%%%%%%%%%%%%%%%%%%%%%%%%%%%%%%%%%%%%%%%%%%%%%%%%%%%%%%%%%%%%%%%%%%%%%%%%%
%%%%%%%%%%%%%%%%%%%%%%%%%%%%%%%%%%%%%%%%%%%%%%%%%%%%%%%%%%%%%%%%%%%%%%%%%%

\par In modern times, the accuracy of spacecraft tracking requires a very detailed modeling of the light propagation in order to compute range and Doppler observables. For example, the Cassini spacecraft reaches the level of few meters accuracy for the range and $3 \times 10^{-6} \,\rm{m/s}$ for the Doppler~\citep{iess:2007ve,bertotti:2003uq,kliore:2004zr} while the future BepiColombo mission should reach an accuracy of $10 \,\rm{cm}$ on the range and $10^{-6}\, \rm{m/s}$ on the Doppler~\citep{milani:2002vn,iess:2009fk}. Similar accuracies are expected for the JUNO mission~\citep{anderson:2004uq}, which shall reach the Jovian system by mid-2016.

\par The computation of radioscience observables as well as the determination of astrometric observables (VLBI tracking \cite{duev:2012fk}) requires determining the propagation of light in a curved space-time.  In this context, several approaches exist.
Assuming that the metric is known, solving the null geodesic equations \cite{san-miguel:2007hc} or the eikonal equation \cite{ashby:2010fk} is the standard method allowing to get all the information about light propagation between two point-events. Many solutions have been proposed in the post-Newtonian~(PN) and in the post-Minkowskian~(PM) approximations when dealing with the bending effects due to the mass multipole moments of the bodies in the Solar System~\cite{klioner:1992ly,kopeikin:1997ys,klioner:2003fk,angonin-willaime:2004kq,crosta:2006rw,kopeikin:2006ly,kopeikin:2007vn,zschocke:2011kx}. On the other hand, the effects of the motion of monopoles on the light propagation have also been studied \cite{klioner:1992ly,kopeikin:1999kl,kopeikin:2002fr,klioner:2003hs,klioner:2003lo}. 
A different approach is also available, initially based on the Synge World Function~\citep{linet:2002ly,le-poncin-lafitte:2004cr,teyssandier:2008fk} and then on the Time Transfer Functions~(TTF)~\citep{teyssandier:2008nx,hees:2014fk}. In this formalism, the computation of the coordinate light time, the frequency shift and the light deflection can be computed as integrals of functions of the components of the metric tensor over a straight line joining the emitter and the receiver of the signal~\citep{teyssandier:2008nx,hees:2014fk}. This method has already been successfully used to compute the propagation of light in different configurations. For example, the TTF in the field of a stationary axisymmetric body has been determined at the first post-Newtonian~(1PN) approximation~\cite{linet:2002ly,le-poncin-lafitte:2008fk}. The light propagation in the field of moving monopoles at 1.5 post-Newtonian order has also been treated~\cite{bertone:2014uq} . Finally, the TTF in the field of a static monopole up to the second and third post-Minkowskian~(3PM) approximation has also been determined~\cite{teyssandier:2012uq,linet:2013fk,teyssandier:2014mz,hees:2014fk}.

\par In this paper, we use the Time Transfer Functions formalism to compute the coordinate propagation time, the frequency shift and the deflection of light in the field of uniformly moving axisymmetric bodies and in the field of arbitrarily moving point masses. In Sec.~\ref{sec:ttf}, we briefly review how the radioscience and astrometric observables can be determined from the TTF and its derivatives. Then, in Sec.~\ref{sec:genmetric}, we determine the space-time metric describing the geometry in the field of a uniformly moving axisymmetric body and we remind the metric describing the field of arbitrarily moving point masses. In Sec.~\ref{sec:ttfmoving}, we use these metrics to develop a general expression of the TTF. A general result is given in the form of an integral computable numerically. Moreover, an analytical result is developed in the case of a uniform motion. The derivatives of the TTF are also determined. In Sec.~\ref{sec:movaxi}, we particularize our results in the case of a uniformly  moving axisymmetric body by determining analytically the contribution of each multipole to the TTF. Finally, in Sec.~\ref{sec:juno}, we apply our results to determine the different relativistic contributions to the radioscience tracking of the JUNO spacecraft in the Jovian system. The contributions of the Sun and Jupiter moving monopoles and of the Jupiter moving $J_2$ are identified. Finally, we give our conclusions and general remarks in Sec.~\ref{sec:conclusion}.

\section{Notation and conventions}  \label{sec:not}
In this paper $c$ is the speed of light in a vacuum and $G$ is the Newtonian gravitational constant. The Lorentzian metric of space-time $V_4$ is denoted by $g$. The signature adopted for $g$ is $(+---)$. We suppose that space-time is covered by some global quasi-Galilean coordinate system $(x^\mu)=(x^0,\bx )$, where $x^0=ct$, $t$ being a time coordinate, and $\bx=(x^i)$. We assume that the curves of equations $x^i$ = const are timelike, which means that $g_{00}>0$ anywhere. We employ the vector notation $\ba$ in order to denote $(a^1,a^2,a^3)=(a^i)$. 
Considering two such quantities $\ba$ and $\bb$ we use $\ba \cdot \bb$ to denote $a^ib^i$ (Einstein convention on repeated indices is used). The quantity $\vert \ba \vert$ stands for the ordinary Euclidean norm of $\ba$. 
For any quantity $f(x^{\lambda})$, $f_{, \alpha}$ denotes the partial derivative of $f$ with respect to $x^{\alpha}$.

%%%%%%%%%%%%%%%%%%%%%%%%%%%%%%%%%%%%%%%	
\section{Time Transfer Function and observables}\label{sec:ttf}
%%%%%%%%%%%%%%%%%%%%%%%%%%%%%%%%%%%%%%%
Let $x_A=(ct_A,\bx_A)$ and $x_B=(ct_B,\bx_B)$ be two events of space-time supposed to be connected by a unique light ray. They denote the emission and reception point of the electromagnetic signal. The coordinate light time of a photon connecting $x_A$ and $x_B$ is given by the TTF~\cite{le-poncin-lafitte:2004cr,jaekel:2005zr,jaekel:2006uq,teyssandier:2008nx} as
\begin{equation}\label{eq:ttf}
	t_B-t_A=\mathcal T(\bx_A,t_B,\bx_B)=\frac{R_{AB}}{c}+\frac{1}{c}\Delta(\bx_A,t_B,\bx_B)
\end{equation}
where $\mathcal T(\bx_A,t_B,\bx_B)$ is the TTF\footnote{In this paper, we used the reception TTF. Similar results can be obtained using the emission TTF which depends on $t_A$ instead of $t_B$ \cite{teyssandier:2008nx}.}, $R_{AB}=\left|\bx_B-\bx_A\right|$ and $\Delta(\bx_A,t_B,\bx_B)$ is the so-called ``delay function"\footnote{In this paper, we call for simplicity $\Delta(\bx_A,t_B,\bx_B)$ a ``delay function" even though it has the dimension of a distance.}.

As developed in details in \cite{hees:2014fk}, the range, Doppler and astrometric observables can all be computed from the TTF. The range is directly related to the coordinate time of flight of the photon  through a coordinate transformation (see also \cite{hees:2012fk}). 

The frequency shift is given by \cite{hees:2012fk,teyssandier:2008fk,teyssandier:2009uq}
\begin{equation}\label{eq:doppler}
	\frac{\nu_B}{\nu_A}= \frac{\left[ g_{00}+2g_{0i} \beta^i +g_{ij} \beta^i\beta^j \right]^{1/2}_A}{\left[ g_{00}+2g_{0i} \beta^i +g_{ij} \beta^i\beta^j \right]^{1/2}_B}\times	 \frac{1-N^i_{AB} \beta^i_B -\beta^i_B \frac{\partial \Delta_r}{\partial x^i_B}-\frac{1}{c}\frac{\partial \Delta_r}{\partial t_B}}{1-N^i_{AB} \beta^i_A + \beta^i_A \frac{\partial \Delta_r}{\partial x^i_A}}
\end{equation}
where $\beta^i_{A/B}=dx^i_{A/B}/cdt$ is the coordinate velocity.

The astrometric observables are directly related to the TTF through the use of \cite{le-poncin-lafitte:2004cr}
\begin{equation} \lb{eq:astrok}
	\left(\hat k_i\right)_B=\left(\frac{k_i}{k_0}\right)_B=- c \, \frac{\partial  {\cal T}_{r}}{\partial x^{i}_{B}} \left[1 - \frac{\partial  {\cal T}_{r}} {\partial t_B}\right]^{-1}=-\left(N^i_{AB}+\frac{\partial \Delta_r}{\partial x^{i}_{B}}\right)\times\left[1 - \frac{1}{c}\frac{\partial  \Delta_{r}} {\partial t_B}\right]^{-1}\,  , 
\end{equation}
where $k_\mu$ are the covariant components of the tangent vector to the photon trajectory $\left(k^\mu\right)_B=\left.dx^\mu/d\lambda\right|_B$ ($\lambda$ being an affine parameter) at $\bx_B$ and $\bN_{AB}=\dfrac{\bR_{AB}}{R_{AB}}=\dfrac{\bx_{B}-\bx_A}{R_{AB}}$ .

Finally, the angular distance between two light rays coming from two different sources can also be related to $(\widehat k_i)_B$ \cite{teyssandier:2006fk,hees:2014fk}.

Therefore, the computation of the TTF (or equivalently of the delay function) and its derivatives is crucial in order to analyze different effects on observations done using light propagation.

%%%%%%%%%%%%%%%%%%%%%%%%%%%%%%%%%%%%%
\section{Metric  at first post-Minkowskian approximation} \label{sec:genmetric}
%%%%%%%%%%%%%%%%%%%%%%%%%%%%%%%%%%%%%
\subsection{Uniformly moving axisymmetric body}\label{sec:metricUnif}
Let us suppose that the gravitational field is generated by an ensemble of axisymmetric bodies. We are interested in calculating the contributions of the mass multipoles and of the motion of the bodies on light propagation. The first step is to consider the metric describing such a space-time. The metric for each of the bodies at 1PM order in its own local reference system is given by $G_{\mu\nu}=\eta_{\mu\nu}+ H_{\mu\nu} $ where $H_{\mu\nu}$ is given by \cite{soffel:2003bd}
\begin{subequations}\label{eq:iaumetricGCRS}
	\begin{eqnarray}
		H_{00}&=&-2\frac{W(X^\alpha)}{c^2}+\mathcal O(G^2) \; , \\
		H_{0i}&=&0  \; , \\
		H_{ab}&=&-2\delta_{ab}\frac{W(X^\alpha)}{c^2}+\mathcal O(G^2) \; ,
	\end{eqnarray}
\end{subequations}
the spin multipoles beeing neglected. Let us stress that the potential $W$ depends on the local coordinate $X^\alpha=(cT,\bX)$.

We can now perform a Poincar\'e transformation in order to obtain the metric in the case of a uniformly moving body. The procedure is similar to what is developed in \cite{zschocke:2014fk}. The coordinate transformation is given by
\begin{equation}\label{eq:transfCoord}
	x^\mu=b^\mu+\Lambda^\mu_\alpha X^\alpha
\end{equation}
where $x^\mu=(ct,\bx)$ are the coordinates of the global reference system and $\Lambda^\mu_\alpha$ is given by
\begin{equation}\label{eq:lambda}
		\Lambda^0_0=\gamma_p \; , \qquad \Lambda^0_i=\Lambda^i_0=\gamma_p\beta^i_p \; , \qquad \Lambda^j_i=\delta_{ij}+\frac{\gamma_p^2}{1+\gamma_p}\beta_p^i\beta_p^j \; ,
\end{equation}
where $\beta_p^i=v_p^i/c$, $v_p^i$ is the coordinate velocity of the body and $\gamma_p={1}/{\sqrt{1-\beta_p^2}}$ with $\beta_p=\left|\bbe_p\right|$. Note that $b^\mu$ is a constant four-vector that specifies the origin of the coordinate system: it points from the origin of the global reference system to the origin of the co-moving frame at $T=0$ \cite{zschocke:2014fk}. We have 
\begin{equation}
	b^i=x^i_p(t_0) \qquad \textrm{and} \qquad b^0=ct_0
\end{equation}
and the trajectory of the moving body in the global frame is given by
\begin{equation} \lb{eq:balpha}
	\bx_p(t)=\bx_{p} (t_0)+c\bbe_p (t-t_0) \; .
\end{equation}
The inverse coordinate transformation is given by
\begin{equation}\label{eq:coordInverse}
	X^\alpha=\tilde\Lambda^\alpha_\mu(x^\mu-b^\mu) \; ,
\end{equation}
where $\tilde \Lambda^\alpha_\mu$ is the inverse of $\Lambda^\mu_\alpha$ and is given by
\begin{equation} \lb{eq:tildelambda}
	\tilde\Lambda^0_0=\gamma_p,\qquad \tilde\Lambda^i_0=\tilde\Lambda^0_i=-\gamma_p\beta^i_p, \qquad \tilde\Lambda^j_i=\delta_{ij}+\frac{\gamma_p^2}{1+\gamma_p}\beta^i_p\beta^j_p.
\end{equation}
The metric transformation is given by
\begin{equation}
	g^{\mu\nu}=\eta^{\mu\nu}+h^{\mu\nu}=\Lambda^\mu_\alpha\Lambda^\nu_\beta G^{\alpha\beta}=\Lambda^\mu_\alpha\Lambda^\nu_\beta \left(\eta^{\alpha\beta}+H^{\alpha\beta}\right) \, ,
\end{equation}
which leads to
\begin{equation} \lb{eq:hmunu}
	h^{\mu\nu}=\Lambda^\mu_\alpha\Lambda^\nu_\beta H^{\alpha\beta}.
\end{equation}
From Eq.~(\ref{eq:iaumetricGCRS}), we have\footnote{Notice that we define $H^{\alpha\beta}=G^{\alpha\beta}-\eta^{\alpha\beta}$ which at the linear order is given by $H^{\alpha\beta}=-\eta^{\alpha\mu}\eta^{\beta\nu}H_{\mu\nu}$.} $H^{\alpha\beta}=\frac{2W}{c^2}\delta^{ab}$. The introduction of this expression and the expression of $\Lambda^\mu_\alpha$ given by Eq.~(\ref{eq:lambda}) into Eq.~\eqref{eq:hmunu} leads to 
\begin{subequations}\label{eq:metric1}
	\begin{eqnarray}
		h^{00}&=&\frac{2W(X^\alpha)}{c^2}\gamma_p^2(1+\beta_p^2) + \mathcal O(G^2)\; ,\\
		h^{0i}&=&\frac{4W(X^\alpha)}{c^2}\beta_p^i\gamma^2_p + \mathcal O(G^2) \; , \\
		h^{ij}&=&\frac{2W(X^\alpha)}{c^2}(\delta_{ij}+2\beta_p^i\beta_p^j\gamma_p^2) + \mathcal O(G^2) .
	\end{eqnarray}
\end{subequations}
It is worth mentioning that this metric is a generalization of the IAU metric~\cite{soffel:2003bd} which can be recovered in the limit of small $\beta_p$. This limit is explicitly developed in Appendix~\ref{app:limitiau1}. Let us also stress that $W$ still depends on the local coordinates $X^\alpha$. Therefore, we still need to use the coordinate transformation~(\ref{eq:coordInverse}) to express the potential $W$ as a function of the global coordinates $x^\alpha$. More precisely, we get
\begin{equation}\label{eq:XLW}
	W=W(X^\alpha)=W(\tilde\Lambda^\alpha_\mu(x^\mu-b^\mu)).
\end{equation}
The metric (\ref{eq:metric1}) describes the geometry generated by a uniformly moving body at 1PM. The metric describing the geometry due to an ensemble of $N$ uniformly moving bodies is then given by
\begin{subequations}\label{eq:metric}
	\begin{eqnarray}
		h^{00}&=&\sum_{p=1}^N \frac{2W_p}{c^2}\gamma_p^2(1+\beta_p^2)+ \mathcal O(G^2)\; ,  \\
		h^{0i}&=&\sum_{p=1}^N\frac{4W_p}{c^2}\beta_p^i\gamma^2_p + \mathcal O(G^2)\; ,\\
		h^{ij}&=&\sum_{p=1}^N \frac{2W_p}{c^2}(\delta_{ij}+2\beta_p^i\beta_p^j\gamma_p^2)+ \mathcal O(G^2)\; .
	\end{eqnarray}
\end{subequations}	
In the case of an axisymmetric body, the Newtonian potential can be decomposed in a multipolar expansion
\begin{equation}\label{eq:multipole}
	W_p(X^i)=\frac{GM_p}{R}\left[1-\sum_{n=2}^\infty J_{np}\left(\frac{r_{pe}}{R}\right)^n P_n\left(\frac{\bk_p\cdot \bX}{R}\right)\right],
\end{equation}
where $\bk_p$ denotes the unit vector along the symmetry axis of the body $p$, $M_p$ is the mass of the body $p$, $J_{np}$ are its mass multipole moments, $P_n$ are the Legendre polynomials, $r_{pe}$ is the equatorial radius of body $p$ and $R=\left| \bX \right|$. In this paper, we assume that the symmetry axis of the body $\bk_p$ is time independent, which means we neglect the precession and nutation of the body.

\subsection{Arbitrarily moving point masses}\label{sec:metricAcc}
The determination of the metric describing the geometry around an arbitrarily moving extended body at the post-Minkowskian approximation is very complex. In particular, one cannot simply use an instantaneous Lorentz transformation  but a local accelerated reference system has to be defined (see the discussion in the conclusion of \cite{zschocke:2014fk}). This is beyond the scope of this paper. Nevertheless, the metric for arbitrarily moving point masses at the post-Minkowskian approximation has been already determined using the Li\'enard-Wiechert potentials \cite{kopeikin:1999kl,kopeikin:2002fr,zschocke:2014fk}. In Appendix~\ref{app:metricAcc}, we briefly remind how to compute this metric.

The space-time metric describing the geometry around an arbitrarily moving point mass can be written as \cite{kopeikin:1999kl,kopeikin:2002fr,zschocke:2014fk} (see also Appendix~\ref{app:metricAcc})
\begin{subequations}\label{eq:metricacc} 
	\begin{eqnarray}
		h^{00}&=&\frac{2W(X^i)}{c^2}\gamma_{pr}^2(1+\beta_{pr}^2) + \mathcal O(G^2)\; ,\\
		h^{0i}&=&\frac{4W(X^i)}{c^2}\beta_{pr}^i\gamma^2_{pr} + \mathcal O(G^2) \; , \\
		h^{ij}&=&\frac{2W(X^i)}{c^2}(\delta_{ij}+2\beta_{pr}^i\beta_{pr}^j\gamma_{pr}^2) + \mathcal O(G^2) .
	\end{eqnarray}
\end{subequations}
with $W(X^i)=GM/R$ (since this metric is only valid for point masses),  
\begin{equation}\label{eq:transfAcc}
	X^i=-\beta^i_{pr}\gamma_{pr}r_{pr}+r^i_{pr}+\frac{\gamma^2_{pr}}{1+\gamma_{pr}}\beta^i_{pr}(\bm \beta_{pr}\cdot \br_{pr})
\end{equation}
and where the index $r$ denotes quantities that have to be evaluated at the retarded time $t_r$ defined by
\begin{equation}\label{eq:retarded}
		t_r=t - \frac{|\bx-\bx_p(t_r)|}{c}=t-\frac{r_{pr}}{c} \; ,
\end{equation}
where $r_{pr}=|\bx-\bx_p(t_r)|$ and $\bx_p(t_r)$ is the position of the body $p$ at the retarded time.
The expression of the potential $W(X^i)$ can then be explicitly written as
\begin{equation}
	W(X^i)=\frac{GM}{|X^i|}=\frac{GM}{\gamma_{pr}\left(r_{pr}-(\br_{pr}\cdot\bm \beta_{pr})\right)}.
\end{equation}
In the limit of small velocities, the expression of the IAU metric is recovered (see Appendix~\ref{app:limitiau2}). Finally the metric for an ensemble of masses is the sum of the metrics generated by each body.

%%%%%%%%%%%%%%%%%%%%%%%%%%%%%%%%%%%%%%%%
\section{Time Transfer Function at generalized 1PM approximation}\label{sec:ttfmoving}
%%%%%%%%%%%%%%%%%%%%%%%%%%%%%%%%%%%%%%%%
In~\cite{teyssandier:2008fk}, a PM expansion of the TTF is presented. It develops the TTF in terms of integrals of functions of the metric components over a straight line between the emitter and the receiver of a light signal. At 1PM order, the delay function is given by~\cite{teyssandier:2009uq} as
\begin{equation}\label{eq:deltaGen}
	\Delta(\bx_A,t_B, \bx_B)=\frac{R_{AB}}{2}\int_0^1 \left[h^{00}-2N^i_{AB}h^{0i}+N^i_{AB}N^j_{AB}h^{ij}\right]_{z^\alpha(\lambda)}d\lambda +\mathcal O(G^2).
\end{equation}
where the integral is taken along a straight line parametrized by
\begin{subequations} \label{eq:zalpha}
	\begin{eqnarray} 
		z^0(\lambda)&=&ct=ct_B-\lambda R_{AB}\label{eq:time}\\
		\bz(\lambda)&=&\bx_B-\lambda \bR_{AB}=\bx_B(1-\lambda)+\lambda \bx_A.
	\end{eqnarray}
\end{subequations}

%%%%%%%%%%%%%%%%%%%%%%%%%%%%%%%%%%%%%%%%%%%
\subsection{General expression in the case of uniform motion}\label{sec:uniform}
%%%%%%%%%%%%%%%%%%%%%%%%%%%%%%%%%%%%%
As it can be seen from the expression of the metric (\ref{eq:metric}), $\Delta$  can be written as a sum of delay functions generated by each individual body $\Delta=\sum_{p=1}^N \Delta_p$. Replacing the expression of the metric (\ref{eq:metric}) in (\ref{eq:deltaGen}) gives
\begin{equation}\label{eq:deltap} 
	\Delta_p(\bx_A,t_B, \bx_B)=\frac{2R_{AB}}{c^2}\int_0^1 \gamma_p^2\left(1-\bN_{AB}\cdot\bbe_p\right)^2 W_p\big(\tilde\Lambda^i_\mu (z^\mu(\lambda)-b^\mu)\big)d\lambda.
\end{equation}
It is useful to express the argument appearing in the expression of the potential $W_p$ in the right hand side of (\ref{eq:deltap}) as
\begin{eqnarray} \lb{eq:tildelambdaiexp}
	\tilde\Lambda^i_\mu (z^\mu(\lambda)-b^\mu)&=&-\beta^i_p\gamma_p c(t-t_0)+z^i(\lambda)-x^i_{p0}+ \frac{\gamma_p^2}{1+\gamma_p}\beta^i_p\ \bbe_p\cdot (\bz(\lambda)-\bx_{p0})\nonumber\\
	&&=x^i_B+\frac{\gamma_p^2}{1+\gamma_p}\beta^i_p\left[\bbe_p\cdot(\bx_B-\bx_{p0})\right]-x^i_{p0}-\gamma_p v^i_p(t_B-t_0)\label{eq:bz}\\
	&&\qquad-\lambda R_{AB}\left[N^i_{AB}-\gamma_p\beta_p^i+\frac{\gamma_p^2}{1+\gamma_p}\beta^i_p(\bbe_p\cdot\bN_{AB})\right]\nonumber \; ,
\end{eqnarray}
by using Eq.~\eqref{eq:balpha}, Eq.~\eqref{eq:tildelambda} and Eq.~\eqref{eq:zalpha} and by posing $\bx_{p0} \equiv \bx_p(t_0)$.  It is also possible to rewrite Eq.~\eqref{eq:tildelambdaiexp} in a more compact form as
\begin{equation}
		\tilde\Lambda^i_\mu (z^\mu(\lambda)-b^\mu)=R^i_{pB}-\lambda G^i_{AB}	
\end{equation}
by setting
\begin{subequations}\label{eq:newVar}
	\begin{eqnarray}
		\bR_{pX}&=&\bx_X+\frac{\gamma_p^2}{1+\gamma_p}\bbe_p\Big[\bbe_p\cdot(\bx_X-\bx_{p0})\Big]-\bx_{p0}-\gamma_p \bv_p(t_B-t_0)\label{eq:Rpb} \; , \\
		\bG_{AB}&=&R_{AB}\left[\bN_{AB}-\gamma_p\bbe_p+\frac{\gamma_p^2}{1+\gamma_p}\bbe_p(\bbe_p\cdot\bN_{AB})\right]=R_{AB}\, \bg_{pAB} \label{eq:Gab}  
	\end{eqnarray}
and
	\begin{equation}
		\bg_{pAB}=\bN_{AB}-\gamma_p\bbe_p+\frac{\gamma_p^2}{1+\gamma_p}\bbe_p(\bbe_p\cdot\bN_{AB}).
	\end{equation}
\end{subequations}
Let us denote by $I$ the integral appearing in the TTF expression (\ref{eq:deltap}) in the case where the body $p$ is static. Therefore $I$ is given by
\begin{equation}\label{eq:I}
	I(\bx_{pA},\bx_{pB})=\tilde I(\bR_{AB},\bx_{pB})=\int_0^1 W_p\big(\bx_{pB}-\lambda \bR_{AB}\big)d\lambda \; ,
\end{equation}
where $\bx_{pA}=\bx_A-\bx_p$ and $\bx_{pB}=\bx_B-\bx_p$.
Usually, the solution of this integral is given in terms of $\bx_{pA}$ and $\bx_{pB}$ but formally, the integral depends on $\bR_{AB}$ and $\bx_{pB}$. The transition between the two expressions of $I$ and $\tilde I$ in the static case is trivial because  $\bx_{pA}=\bx_{pB}-\bR_{AB}$. However, this transition no longer applies in the moving case and it has to be replaced in Eq.~(\ref{eq:deltap}) by
\begin{equation} \label{eq:Imov}
	\tilde I(\bG_{AB},\bR_{pB})=\int_0^1W_p\big(\bR_{pB}-\lambda \bG_{AB}\big) d\lambda \;
\end{equation}
with the two variables defined by Eqs.~(\ref{eq:newVar}) and similarly to what proposed in~\citep{bertone:2014uq}.

Therefore, all the results in the moving case can be derived from the expressions used in the static case by replacing $\bx_{pB}$ by $\bR_{pB}$ (\ref{eq:Rpb}) and $\bR_{AB}$ by $\bG_{AB}$ (\ref{eq:Gab}). We can use the conversions given below, where for each ``static case" quantity on the left we give the ``moving case" equivalent on the right. We get
\begin{subequations}\label{eq:substitution}
	\begin{eqnarray}
		\bx_{pB} &\rightarrow& \bR_{pB}=\bx_B+\frac{\gamma_p^2}{1+\gamma_p}\bbe_p\left[\bbe_p\cdot(\bx_B-\bx_{p0})\right]-\bx_{p0}-\gamma_p \bv_p(t_B-t_0) \; , \\
		r_{pB}=\left|\bx_{pB}\right|&\rightarrow& R_{pB}=\left|\bR_{pB}\right| \; ,\\
		\bn_{pB} &\rightarrow& \bN_{pB}=\frac{\bR_{pB}}{R_{pB}} \; ,\\
		\bR_{AB} &\rightarrow& \bG_{AB}=R_{AB}\bg_{pAB}=R_{AB}\left[\bN_{AB}-\gamma_p\bbe_p+\frac{\gamma_p^2}{1+\gamma_p}\bbe_p(\bbe_p\cdot\bN_{AB})\right] \; ,\\
		R_{AB}&\rightarrow& R_{AB}\gamma_p(1-\bbe_p\cdot\bN_{AB} ) \; , \\
		\bN_{AB}&\rightarrow &\frac{\bg_{pAB}}{g_{pAB}}=\frac{\bg_{pAB}}{\gamma_p(1-\bbe_p\cdot\bN_{AB})} \; ,\\
		\bx_{pA}=\bx_{pB}-\bR_{AB}&\rightarrow &\bR_{pB}-\bG_{AB}=\bR_{pA}+\gamma_p\bm\beta_{p} R_{AB}  \; ,\\
		r_{pA}=\left|\bx_{pA}\right|&\rightarrow& R_{pA}=\left|\bR_{pA}\right| \; , \\
		\bn_{PA} &\rightarrow& \frac{\bR_{pA}+\gamma_p\bm\beta_{p} R_{AB}  }{\left|\bR_{pA}+\gamma_p\bm\beta_{p} R_{AB}  \right|} \; ,
	\end{eqnarray}
\end{subequations}
with
\begin{equation}
	g_{pAB}=\left|\bg_{pAB}\right|=\gamma_{p}\left(1-\bbe_p\cdot\bN_{AB}\right) \; ,
\end{equation} 
and $\bm R_{pX}$ given by Eq.~\eqref{eq:Rpb}.
Therefore, we can rewrite Eq.~\eqref{eq:deltap} as
\begin{equation}\label{eq:deltapbis} 
	\Delta_p(\bx_A,t_B, \bx_B)=\frac{2R_{AB}}{c^2} \gamma_p^2\left(1-\bN_{AB}\cdot\bbe_p\right)^2 \int_0^1 W_p\big( R^i_{pB} - \lambda G^i_{AB} \big)d\lambda .
\end{equation}
Then, using the definition of $I$ from Eqs.~(\ref{eq:I})-(\ref{eq:Imov}) and the correspondences (\ref{eq:substitution}), we are able to express the exact form of the TTF in the field of moving bodies as
\begin{equation}\label{eq:deltap2}
		\Delta_p(\bx_A,t_B,\bx_B)=\frac{2R_{AB}}{c^2}\gamma_p^2\left(1-\bN_{AB}\cdot\bbe_p\right)^2I(\bR_{pA}+\gamma_p\bm\beta_{p} R_{AB},\bR_{pB}) \; ,
\end{equation}
with $\bR_{pX}$ given by (\ref{eq:Rpb}). The last expression can also be written as
\begin{eqnarray}
		\Delta_p(\bx_A,t_B,\bx_B)&=&\frac{\gamma_p^2\left(1-\bN_{AB}\cdot\bbe_p\right)^2}{g_{pAB}}\tilde\Delta_{p}(\bR_{pA}+\gamma_p\bm\beta_{p} R_{AB},\bR_{pB})\\
		&=&\gamma_p\left(1-\bN_{AB}\cdot\bbe_p\right)\tilde\Delta_{p}(\bR_{pA}+\gamma_p\bm\beta_{p} R_{AB},\bR_{pB})\label{eq:deltap3} \; ,
\end{eqnarray}
where $\tilde\Delta(\bx_{pA},\bx_{pB})$ is the expression of the static TTF. This particularly simple equation is very useful since it allows one to determine the TTF of a uniformly moving body from the corresponding static TTF.

The derivatives of the TTF, needed to compute the frequency shift~\eqref{eq:doppler} and the astrometric direction~\eqref{eq:astrok}, can be computed from (\ref{eq:deltap3}) remembering Eq.~\eqref{eq:newVar}. In the case of a uniformly moving body, their expressions are given by
\begin{subequations}\label{eq:derdeltap2}
	\begin{eqnarray}
		\frac{\partial \Delta_p(\bx_A,t_B,\bx_B)}{\partial x^i_A}&=&\gamma_p\left(1-\bN_{AB}\cdot\bbe_p\right)\tilde\Delta_{p,jA}(\bR_{pA}+\gamma_p\bm\beta_{p} R_{AB},\bR_{pB})\left[\delta_{ij}-\gamma_p\beta_p^j\left(N^i_{AB}-\frac{\gamma_p}{1+\gamma_p}\beta^i_p\right)\right]\nonumber\\
		&&+\gamma_p\frac{\beta^i_p-N^i_{AB}\bbe_p\cdot\bN_{AB}}{R_{AB}}\tilde \Delta_p(\bR_{pA}+\gamma_p\bm\beta_{p} R_{AB},\bR_{pB}) \; , \\
		\frac{\partial \Delta_p(\bx_A,t_B,\bx_B)}{\partial x^i_B}&=&\gamma_p\left(1-\bN_{AB}\cdot\bbe_p\right)\Bigg\{\tilde\Delta_{p,jB}(\bR_{pA}+\gamma_p\bm\beta_{p} R_{AB},\bR_{pB})\left[\delta_{ij}+\frac{\gamma_p^2}{1+\gamma_p}\beta^i_p\beta^j_p\right] \nonumber\\
		&&\qquad \qquad +\gamma_p\beta_p^jN^i_{AB} \tilde\Delta_{p,jA}(\bR_{pA}+\gamma_p\bm\beta_{p} R_{AB},\bR_{pB})\Bigg\}\\
		&&-\gamma_p\frac{\beta^i_p-N^i_{AB}\bbe_p\cdot\bN_{AB}}{R_{AB}}\tilde \Delta_p(\bR_{pA}+\gamma_p\bm\beta_{p} R_{AB},\bR_{pB})\nonumber \; , \\
	\frac{\partial \Delta_p(\bx_A,t_B,\bx_B)}{\partial t_B}&=&-c \gamma_p^2\left(1-\bN_{AB}\cdot\bbe_p\right) \beta^i_{p}\Big[\tilde\Delta_{p,iA}(\bR_{pA}+\gamma_p\bm\beta_{p} R_{AB},\bR_{pB}) \nonumber  \\
                  && \qquad \qquad + \tilde\Delta_{p,iB}(\bR_{pA}+\gamma_p\bm\beta_{p} R_{AB},\bR_{pB})\Big] \; ,
	\end{eqnarray}
\end{subequations}
where $\tilde\Delta_{p,iX}(\bx_A,\bx_B)$ is the expression of the derivative of the static TTF with respect to $\bx_X$
\begin{equation}
	\tilde\Delta_{p,iX}(\bx_A,\bx_B)=\frac{\partial \tilde\Delta_{p}(\bx_A,\bx_B)}{\partial x^i_X} \; .
\end{equation} 
It is worth mentioning that in the static case, we have the relation
\begin{equation}
	\tilde \Delta_p(\bx_A,\bx_B)=	\tilde \Delta_p(\bx_B,\bx_A)
\end{equation}
and consequently
\begin{equation}
	\tilde \Delta_{p,iB}(\bx_A,\bx_B)=	\tilde \Delta_{p,iA}(\bx_B,\bx_A).
\end{equation}
Therefore, the expression of the derivatives of the TTF in the ``moving case" is also obtained by inserting into Eqs.~(\ref{eq:derdeltap2}) the static TTF and its derivatives.. We present an application in the field of moving axisymmetric bodies in Section~\ref{sec:movaxi}.

%%%%%%%%%%%%%%%%%%%%%%%%%%%%%%%%%
\subsection{Case of a non-uniform motion}\label{sec:genMotion}
%%%%%%%%%%%%%%%%%%%%%%%%%%%%%%%%%
The previous section gives the exact solution of the TTF in the field of uniformly moving bodies. If the bodies undergo acceleration, it is still possible to use the previous formula, which corresponds to neglect higher order terms related to the acceleration of the body. In this case, the choice of the parameter $t_0$ introduced in Eq.~(\ref{eq:bz}) becomes critical.
%\begin{equation}
%	\bz(\lambda)-\bx_p(t)=\bx_B-\lambda R_{AB}\bN_{AB}-\bx_{pc}-(t_B-t_c)\bv_{pc}-(t-t_B)\bv_{pc}-\frac{1}{2}(t-t_c)^2\ba_{pc}
%\end{equation}
%where $\ba_{pc}=d\bv(t_c)/dt$ is the acceleration of the body. 
It has been shown \cite{klioner:1992ly,klioner:2003fk,klioner:2003lo} that a good choice of $t_0$ ($i.e.$ which minimizes the approximation error) is given by the time of closest approach of the photon with respect to the body which is given by %\cite{klioner:1992ly,klioner:2003fk,klioner:2003lo}
\begin{equation} \lb{eq:t0}
 t_0= \textrm{max}\left(t_A,t_B- \textrm{max}\left(0,\frac{\bg \cdot (\bx_B-\bx_p(t_B))}{c|\bg|^2}\right)\right)
\end{equation}
with $\bg=\bN_{AB}-\bbe_p(t_B)$.

In the case of arbitrarily moving point masses, it is possible to numerically integrate the TTF (\ref{eq:deltaGen}) using the metric (\ref{eq:metricacc}). This approach has the convenience to be strictly valid at the 1PM order whatever the motion of the bodies. Inserting (\ref{eq:metricacc}) in the expression (\ref{eq:deltaGen}) gives
\begin{eqnarray} \lb{eq:Deltanumeric}
 \Delta_p(\bx_A,t_B,\bx_B)&=&\frac{2R_{AB}}{c^2}\int_0^1 \gamma_{pr}^2 \left(1-\bN_{AB}\cdot\bm\beta_{pr}\right)^2 \times \\
&&\qquad W_p\left(z^i(\lambda)-x^i_p(t_r)+\frac{\gamma_{pr}^2}{1+\gamma_{pr}}\beta^i_{pr} \bbe_{pr}\cdot\left(\bz(\lambda)-\bx_{p}(t_r)\right)-c\beta^i_{pr}\gamma_{pr}(t-t_r)\right)d\lambda \; , \nonumber
\end{eqnarray}
where $\gamma_{pr}$ and $\beta_{pr}$ depend on the retarded time coordinate $t_r$ that is related to $t$ through (\ref{eq:retarded}). The integral in Eq.~\eqref{eq:Deltanumeric} can then be evaluated numerically whatever the motion of the body $\bx_p(t)$.

%%%%%%%%%%%%%%%%%%%%%
\subsection{Moving emitter}
In the previous sections, we handle the case where the source of the gravitational field is moving. In general, the emitter and the receiver of the electromagnetic signal are also moving. In this case, the determination of the time transfer requires solving Eq.~(\ref{eq:ttf}), which is now implicit
$$
t_B-t_A=\mathcal T(\bx_A(t_A),t_B,\bx_B)=\frac{\left|\bx_B-\bx_A( t_A)\right|}{c}+\frac{1}{c}\Delta(\bx_A(t_A),t_B,\bx_B).
$$
In practice, the solution of this implicit equation can be determined by an iterative procedure to find $t_A$ (for example, see Eq.~(7) of \cite{hees:2014fk}). Another solution consists in a post-Newtonian expansion of  $t_A$ from the TTF (for example, see Eq.~(6) of \cite{hees:2014fk}). Let us denote by $\bar t_A$ the coordinate time of emission solution of
\begin{equation}
 t_B-\bar t_A= \frac{\left|\bx_B-\bx_A(\bar t_A)\right|}{c} \; ,
\end{equation}
then we can write at first order in $\bar {\bm\beta}_A=\bv_A(\bar t_A)/c$
\begin{equation}\label{eq:sagnac}
t_B-t_A=\frac{\bar R_{AB}}{c}+\Delta(\bx_A(\bar t_A),t_B,\bx_B) - \frac{\bar R_{AB}}{c}\bar\beta_A^i \frac{\partial \Delta(\bx_A(\bar t_A),t_B,\bx_B)}{\partial x^i_A}
\end{equation}
where the ``bar" denotes quantities evaluated at $\bar t_A$ like $\bar R_{AB}=\left|\bx_B-\bx_A(\bar t_A)\right|$. The contribution proportional to $\bar{\bm \beta}_A$ is also known as a Sagnac term. It has the same form as the contribution from the velocity of the source of the gravitational field at first post-Newtonian order as can be seen from Eq.~(\ref{eq:deltap_pn}). The order of magnitude of this contribution can reach a few meters for a JUNO-Earth signal as it can be seen from Fig.~\ref{fig:juno4}. Therefore, when solving iteratively the light-time equation, one needs to include the relativistic perturbations or to take into account the Sagnac terms to avoid the risk of significant errors.

%%%%%%%%%%%%%%%%%%%%%%%%%%%%%%%%%%%%%
\section{Case of uniformly moving axisymmetric bodies} \label{sec:movaxi}
%%%%%%%%%%%%%%%%%%%%%%%%%%%%%%%%%%%%%
We can now use the general procedure presented in the previous section in the case of uniformly moving axisymmetric bodies whose potential is given by the multipole expansion (\ref{eq:multipole}). The TTF in the case of a static axisymmetric body has been computed in~\cite{le-poncin-lafitte:2008fk} and is given by
\begin{equation} \lb{eq:DeltaMpJnp}
	\Delta_p(\bx_{pA},\bx_{pB}) = \Delta_{Mp}(\bx_{pA},\bx_{pB}) + \Delta_{Jpn}(\bx_{pA},\bx_{pB}) \;
\end{equation}
where $\Delta_{Mp}$ represents the mass monopole contribution and $\Delta_{Jpn}$ represents the mass multipoles contribution.

The TTF corresponding to a static monopole is well known~\citep{linet:2002ly} and is given by
\begin{equation}\label{eq:deltamono}
	\tilde\Delta_{Mp}(\bx_{pA},\bx_{pB})=2\frac{GM_p}{c^2}\ln \frac{r_{pA}+r_{pB}+R_{AB}}{r_{pA}+r_{pB}-R_{AB}}.
\end{equation}
By inserting  (\ref{eq:deltamono}) into (\ref{eq:deltap3}) and using the substitutions~(\ref{eq:substitution}), we obtain the TTF in the field of monopoles in uniform motion as
\begin{equation}\label{eq:deltamonomov}
	\Delta_M(\bx_A,t_B,\bx_B)=2\frac{GM_p}{c^2}\gamma_p\left(1-\bN_{AB}\cdot\bbe_p\right) \ln \frac{\left|\bR_{pA}+\gamma_p\bbe_{p} R_{AB}\right|+R_{pB}+\gamma_p R_{AB}(1-\bbe_p\cdot\bN_{AB})}{\left|\bR_{pA}+\gamma_p\bbe_{p} R_{AB}\right|+R_{pB}-\gamma_p R_{AB}(1-\bbe_p\cdot\bN_{AB})} \; ,
\end{equation}
with $\bR_{pX}$ given by (\ref{eq:Rpb}).
On the other hand, the mass multipole contribution $\Delta_{Jpn}$ has been computed in~\cite{le-poncin-lafitte:2008fk} as
\begin{subequations}\label{eq:deltamultipole}
	\begin{eqnarray}
	\tilde\Delta_{J_{np}}(\bx_{pA},\bx_{pB})&=&K_{pn} \sum_{m=1}^{n}\left[\frac{1}{(r_{pA}+r_{pB}-R_{AB})^{n-m+1}} - \frac{1}{(r_{pA}+r_{pB}+R_{AB})^{n-m+1}}\right]\Theta_{nm}(\bx_{pA},\bx_{pB}) \; ,
	\end{eqnarray}
	with $K_{pn} \equiv (1+\gamma)GM_p J_{np}r_{pe}^n/c^2$ and
	\begin{equation}
		\Theta_{nm}(\bx_{pA},\bx_{pB})=(-1)^{n-m}\sum_{i_1,\dots,i_m} '\frac{(n-m)!}{i_1! i_2!\dots i_m!} \prod_{l=1}^m [S_l(\bx_{pA},\bx_{pB})]^{i_l},
	\end{equation}
where the sum $\sum_{i_1,\dots,i_m} '$ denotes the summation over the sets of nonnegative integers $i_1, i_2, \dots, i_m$ satisfying the pair of equations
\begin{equation}\label{eq:integers}
	\begin{cases}
		i_1 +2 i_2 + 3i_3 +\dots + m i_m=n \\
		i_1+i_2+\dots+i_m=n-m+1
	\end{cases}
\end{equation}
and where $S_l(\bx_{pA},\bx_{pB})$ is defined by
	\begin{equation}
		S_l(\bx_{pA},\bx_{pB})=\frac{1}{r_{pA}^{l-1}}C_l^{(-1/2)}\left(\frac{\bk_p\cdot \bx_{pA}}{r_{pA}}\right)+\frac{1}{r_{pB}^{l-1}}C_l^{(-1/2)}\left(\frac{\bk_p\cdot \bx_{pB}}{r_{pB}}\right)
	\end{equation}
	with $C_l^{(-1/2)}(x)$ the Gegenbauer polynomial of degree $l$ and of parameter $-1/2$.
\end{subequations}

Therefore, the multipole term of the TTF for the case of moving axisymmetric bodies is given by inserting (\ref{eq:deltamultipole}) into the relation (\ref{eq:deltap3}) and using the substitutions~(\ref{eq:substitution})
\begin{eqnarray}
	\Delta_{J_{np}}(\bx_A,t_B,\bx_B)&=&\frac{2GM_p J_{np}r_{pe}^n}{c^2}\gamma_p\left(1-\bN_{AB}\cdot\bbe_p\right) \nonumber\\
	&&\times\sum_{m=1}^{n}\left[\frac{1}{(\left|\bR_{pA}+\gamma_p\bbe_p R_{AB}\right|+R_{pB}-R_{AB}\gamma_p(1-\bbe_p\cdot\bN_{AB}))^{n-m+1}} \right.\\
	&& \left.- \frac{1}{(\left|\bR_{pA}+\gamma_p\bbe_p R_{AB}\right|+R_{pB}+R_{AB}\gamma_p(1-\bbe_p\cdot\bN_{AB}))^{n-m+1}}\right]\Theta_{nm}(\bR_{pA}+\bbe_pR_{AB},\bR_{pB})\nonumber
\end{eqnarray}
with $\bR_{pX}$ given by (\ref{eq:Rpb}).

In order to compute the derivatives of the TTF in the case of moving bodies from Eq.~\eqref{eq:derdeltap2}, one also needs the derivatives of the TTF in the static case. The derivative of the TTF in the case of a static monopole is known (see for example \cite{blanchet:2001ud}) and it is given by
\begin{subequations}\label{eq:derdeltamono}
	\begin{eqnarray}
			\tilde\Delta_{Mp,iA}(\bx_{pA},\bx_{pB})&=&-\frac{4GM_p}{c^2}\frac{N^i_{AB}(r_{pA}+r_{pB})+R_{AB}n^i_{pA}}{(r_{pA}+r_{pB})^2-R_{AB}^2}\; ,\\					\tilde\Delta_{Mp,iB}(\bx_{pA},\bx_{pB})&=&+\frac{4GM_p}{c^2}\frac{N^i_{AB}(r_{pA}+r_{pB})-R_{AB}n^i_{pB}}{(r_{pA}+r_{pB})^2-R_{AB}^2}=\tilde\Delta_{p,iA}(\bx_{pB},\bx_{pA}) \; .
	\end{eqnarray}
\end{subequations}
Also, the derivatives of Eq.~(\ref{eq:deltamultipole}) can be computed as
\begin{subequations}\label{eq:derIMult}
	\begin{eqnarray}
\tilde \Delta_{Jpn,iA}(\bx_{pA},\bx_{pB})&=&K_{pn}\sum_{m=1}^n \left\{-(n-m+1)\left[\frac{\bn_{pA}+\bN_{AB}}{(r_{pA}+r_{pB}-R_{AB})^{n-m+2}}-\frac{\bn_{pA}-\bN_{AB}}{(r_{pA}+r_{pB}+R_{AB})^{n-m+2}}\right]\Theta(\bx_{pA},\bx_{pB})\right. \nonumber\\
		&&\left. +\left[\frac{1}{(r_{pA}+r_{pB}-R_{AB})^{n-m+1}} - \frac{1}{(r_{pA}+r_{pB}+R_{AB})^{n-m+1}}\right]\bm\Upsilon_{A|nm}(\bx_{pA},\bx_{pB})\right\},\label{eq:dIdxa}\\
\tilde \Delta_{Jpn,iB}(\bx_{pA},\bx_{pB})&=&K_{pn}\sum_{m=1}^n \left\{-(n-m+1)\left[\frac{\bn_{pB}-\bN_{AB}}{(r_{pA}+r_{pB}-R_{AB})^{n-m+2}}-\frac{\bn_{pB}+\bN_{AB}}{(r_{pA}+r_{pB}+R_{AB})^{n-m+2}}\right]\Theta(\bx_{pA},\bx_{pB})\right.\nonumber\\
		&&\left. +\left[\frac{1}{(r_{pA}+r_{pB}-R_{AB})^{n-m+1}} - \frac{1}{(r_{pA}+r_{pB}+R_{AB})^{n-m+1}}\right]\bm\Upsilon_{B|nm}(\bx_{pA},\bx_{pB})\right\},
	\end{eqnarray}
	where
	\begin{eqnarray}
\bm\Upsilon_{X|nm}(\bx_{pA},\bx_{pB})&=&(-1)^{n-m}\sum_{i_1,\dots,i_m} '\frac{(n-m)!}{i_1! i_2!\dots i_m!}\sum_{l=1}^m i_l[S_l(\bx_{pA},\bx_{pB})]^{i_l-1}\\
&&\qquad \prod_{q=1,q\neq l}^m[S_q(\bx_{pA},\bx_{pB})]^{i_q}\frac{[P_{l-1}(\bk_p\cdot \bn_{pX})\bk_p-P_{l}(\bk_p\cdot \bn_{pX})\bn_{pX}]}{r_{pX}^l}.\nonumber
	\end{eqnarray}
	\end{subequations}
The derivatives of the TTF function in the case of a moving axisymmetric body is then given by combining Eq.~\eqref{eq:deltamono} and Eq.~(\ref{eq:derIMult}) into Eq.~\eqref{eq:DeltaMpJnp} and by using it together with Eq.~\eqref{eq:derdeltamono} and Eq.~(\ref{eq:derIMult}) into Eqs.~(\ref{eq:derdeltap2}) (using the correspondances (\ref{eq:substitution})). 

%%%%%%%%%%%%%%%%%%%%%%%%%%%
\subsection{Particular case: post-Newtonian expansion} \lb{sec:PNexp}
The  section~\ref{sec:uniform} gives a way to compute the TTF in the field of uniformly moving bodies. The obtained expressions are exact at any order in $\bm\beta_{p}$. Nevertheless, a post-Newtonian expression can sometimes be more practical to use in the case of slowly moving bodies. Therefore, we present here an expansion of the previous results in terms of the small parameter $\bbe_p$. An expansion of (\ref{eq:deltap2}) gives
\begin{eqnarray}\label{eq:deltap_pn}
		\Delta_p(\bx_A,t_B,\bx_B)&=&(1-\bm \beta_{p}\cdot \bN_{AB})\tilde\Delta_p(\bx_{pA},\bx_{pB})+\left(R_{AB}-c(t_B-t_0)\right)\beta^i_{p} \tilde\Delta_{p,iA} (\bx_{pA},\bx_{pB}) \\
		&&\qquad-c(t_B-t_0)\beta^i_p\tilde\Delta_{p,iB}(\bx_{pA},\bx_{pB}) \; , \nonumber
\end{eqnarray}
with $\bx_{pX}=\bx_X-\bx_p(t_0)$.
For example, the use of this formula in the case of the moving monopoles leads to
\begin{eqnarray}
	\Delta_p(\bx_A,t_B,\bx_B)&=&2\frac{GM_p}{c^2}\left(1-\bm\beta_{p}\cdot\bN_{AB}\right)\ln \frac{r_{pA}+r_{pB}+R_{AB}}{r_{pA}+r_{pB}-R_{AB}}\nonumber \\
	&&-4 \frac{GM_pR_{AB}}{c^2}\frac{(r_{pA}+r_{pB})\bN_{AB}\cdot\bbe_{p}+R_{AB}\bn_{pA}\cdot\bbe_{p}}{(r_{pA}+r_{pB})^2-R^2_{AB}} \\
	&& +4 \frac{GM_pR_{AB}}{c}(t_b-t_0) \frac{\bbe_p\cdot(\bn_{pA}+\bn_{pB})}{(r_{pA}+r_{pB})^2-R^2_{AB}}+\mathcal O(c^{-4}) \; , \nonumber
\end{eqnarray}
with $\bn_{pX}=\frac{\bx_{pX}}{r_{pX}}=\frac{\bx_{pX}}{\left|\bx_{pX}\right|}$. This expression is equivalent to the one given by Eq.~(20) of \cite{bertone:2014uq}. To obtain this result in such a straightforward way, illustrates the effectiveness of the TTF approach.  

%%%%%%%%%%%%%%%%%%%%%%%%%%%%%%
\subsection{Particular case: the quadrupolar term}
An explicit calculation for each of the multipoles is straightforward given the above formulas. As an example, let us develop explicitly the expression for the quadrupolar term $J_2$. The only sets of integers solutions to Eqs.~(\ref{eq:integers}) are $i_1=2$ for $m=1$ and $\{i_1=0,i_2=1\}$ for $m=2$. As shown in~\cite{le-poncin-lafitte:2008fk}, we obtain
\begin{eqnarray} \label{eq:J2stat}
	\tilde\Delta_{J_{p2}}(\bx_{pA},\bx_{pB})&=&\frac{GM_p}{c^2}\frac{J_{p2}r_{pe}^2}{r_{pA}r_{pB}}\frac{R_{AB}}{1+\bn_{pA}\cdot\bn_{pB}}\times\left[\frac{1-(\bk_p\cdot\bn_{pA})^2}{r_{pA}} + \frac{1-(\bk_p\cdot\bn_{pB})^2}{r_{pB}} \right.\\
	&& \qquad \qquad\left. -\left(\frac{1}{r_{pA}}+\frac{1}{r_{pB}}\right)\frac{\left[\bk_p\cdot(\bn_{pA}+\bn_{pB})\right]^2}{1+\bn_{pA}\cdot\bn_{pB}}\right]\nonumber.
\end{eqnarray}
Therefore, inserting (\ref{eq:J2stat}) into (\ref{eq:deltap3}) and using the substitutions (\ref{eq:substitution}), we obtain
\begin{eqnarray} 
	\Delta_{J_{p2}}(\bx_{pA},\bx_{pB})&=&\frac{GM_p}{c^2} \gamma_p^2\left(1-\bN_{AB}\cdot\bbe_p\right)^2 \frac{J_{p2}r_{pe}^2}{\left|\bR_{pA}+\gamma_p\bm\beta_{p} R_{AB}  \right|R_{pB}}\frac{R_{AB}}{1+\bm{\tilde N_A}\cdot\bN_{pB}}\label{eq:J2mov}\\
	&&\times\left[ \frac{1-(\bk_p\cdot\bm{\tilde N_A} )^2}{\left|\bR_{pA}+\gamma_p\bm\beta_{p} R_{AB}  \right|} + \frac{1-(\bk_p\cdot\bN_{pB})^2}{R_{pB}}   -\left(\frac{1}{\left|\bR_{pA}+\gamma_p\bm\beta_{p} R_{AB}  \right|}+\frac{1}{R_{pB}}\right)\frac{\left[\bk_p\cdot(\bm{\tilde N_A}+\bN_{pB})\right]^2}{1+\bm{\tilde N_A}\cdot\bN_{pB}}\right]\nonumber 
\end{eqnarray}
and
\begin{equation}
	\bm{\tilde N_A}=\frac{\bR_{pA}+\gamma_p\bm\beta_{p} R_{AB}  }{\left|\bR_{pA}+\gamma_p\bm\beta_{p} R_{AB}  \right|} \; .
\end{equation}
The derivative of (\ref{eq:J2stat}) with respect to $\bx_{pA}$ can be computed using Eq.~(\ref{eq:dIdxa}) and is given by
\begin{subequations}\label{eq:DIJ2stat}
\begin{eqnarray} 
\tilde\Delta_{J_{p2},iA}(\bx_{pA},\bx_{pB})&=&	2\frac{GM_p}{c^2}J_{p2}r_{pe}^2\left\{ [\bk_p\cdot(\bn_{pA}+\bn_{pB})]^2\left[\frac{\bn_{pA}+\bN_{AB}}{(r_{pA}+r_{pB}-R_{AB})^3} - \frac{\bn_{pA}-\bN_{AB}}{(r_{pA}+r_{pB}+R_{AB})^3}\right]\right.\nonumber\\
&&-\frac{1}{2}\left[\frac{1-(\bk_p\cdot\bn_{pA})^2}{r_{pA}} + \frac{1-(\bk_p\cdot\bn_{pB})^2}{r_{pB}}\right] \left[\frac{\bn_{pA}+\bN_{AB}}{(r_{pA}+r_{pB}-R_{AB})^2} - \frac{\bn_{pA}-\bN_{AB}}{(r_{pA}+r_{pB}+R_{AB})^2} \right] \nonumber\\ 
&&-\frac{1}{r^3_{pA}}\frac{R_{AB}(r_{pA}+r_{pB})}{r_{pB}^2}\frac{\bk_p\cdot(\bn_{pA}+\bn_{pB})}{(1+\bn_{pA}\cdot\bn_{pB})^2}[\bk_p-(\bk_p\cdot\bn_{pA})\bn_{pA}]\\
&&-\frac{1}{2r^3_{pA}}\frac{R_{AB}}{r_{pB}} \frac{2(\bk_p\cdot\bn_{pA})\bk_p+[1-3(\bk_p\cdot\bn_{pA})^2]\bn_{pA}}{1+\bn_{pA}\cdot\bn_{pB}}\nonumber.
\end{eqnarray}
while the derivatives with respect to $\bx_{pB}$ can be obtained by symmetry as
\begin{equation}
	\tilde\Delta_{J_{p2},iB}(\bx_{pA},\bx_{pB})=\tilde\Delta_{J_{p2},iA}(\bx_{pB},\bx_{pA}) \; .
\end{equation}
\end{subequations}
In order to evaluate the contribution of the moving quadrupole to the derivatives of the time transfer, it is then sufficient to combine Eq.~\eqref{eq:DIJ2stat} and Eq.~\eqref{eq:J2stat} as shown in Eq.~(\ref{eq:derdeltap2}).

%%%%%%%%%%%%%%%%%%%%%%%%%%%%%%%%%%%%%%%%%
\section{Application to JUNO}\label{sec:juno}
%%%%%%%%%%%%%%%%%%%%%%%%%%%%%%%%%%%%%%%%%
As an example, we use the equations presented in previous sections to give estimates of the relativistic corrections on the observables for the JUNO mission. JUNO is currently on his way to Jupiter that will be reached in 2016. The spacecraft will orbit Jupiter during one year. Some of the relativistic perturbations on JUNO orbit have been studied in \cite{iorio:2010kx,iorio:2013yq}. The main goal  of this section is to assess the order of magnitude produced by different effects due to the Sun and Jupiter on the time transfer. We shall use the nominal orbit of the mission around Jupiter obtained using the Naif SPICE toolkit \cite{acton:1996fk} and kernels as well as the DE430 planetary ephemeris \cite{folkner:2014uq}. The expected accuracy for JUNO is of the order of 10 cm on the range and $10^{-6} \, m/s$ on the Doppler \citep{anderson:2004uq}. In the following, we present different relativistic contributions to the 2-ways coordinate light time between Earth and JUNO and the corresponding range-rate. The range-rate has been computed with an integration time of 10 seconds.

In the following figures, all the time scales are given in terms of the coordinate time, which is similar to the TCB introduced in the IAU conventions (see \cite{soffel:2003bd}). The observations, done in terms of local time, can be derived by a relativistic coordinate transformations, which is conventional \cite{soffel:2003bd,moyer:2000uq}. Nevertheless, this transformation will not significantly change the figures presented below.

Fig.~\ref{fig:juno1} represents the lower order time transfer and range rate between JUNO and Earth as well as the relativistic Shapiro correction from the Sun. These corrections are standard. 
\begin{figure}[hbt]
\begin{center}
\includegraphics[width=0.48\textwidth]{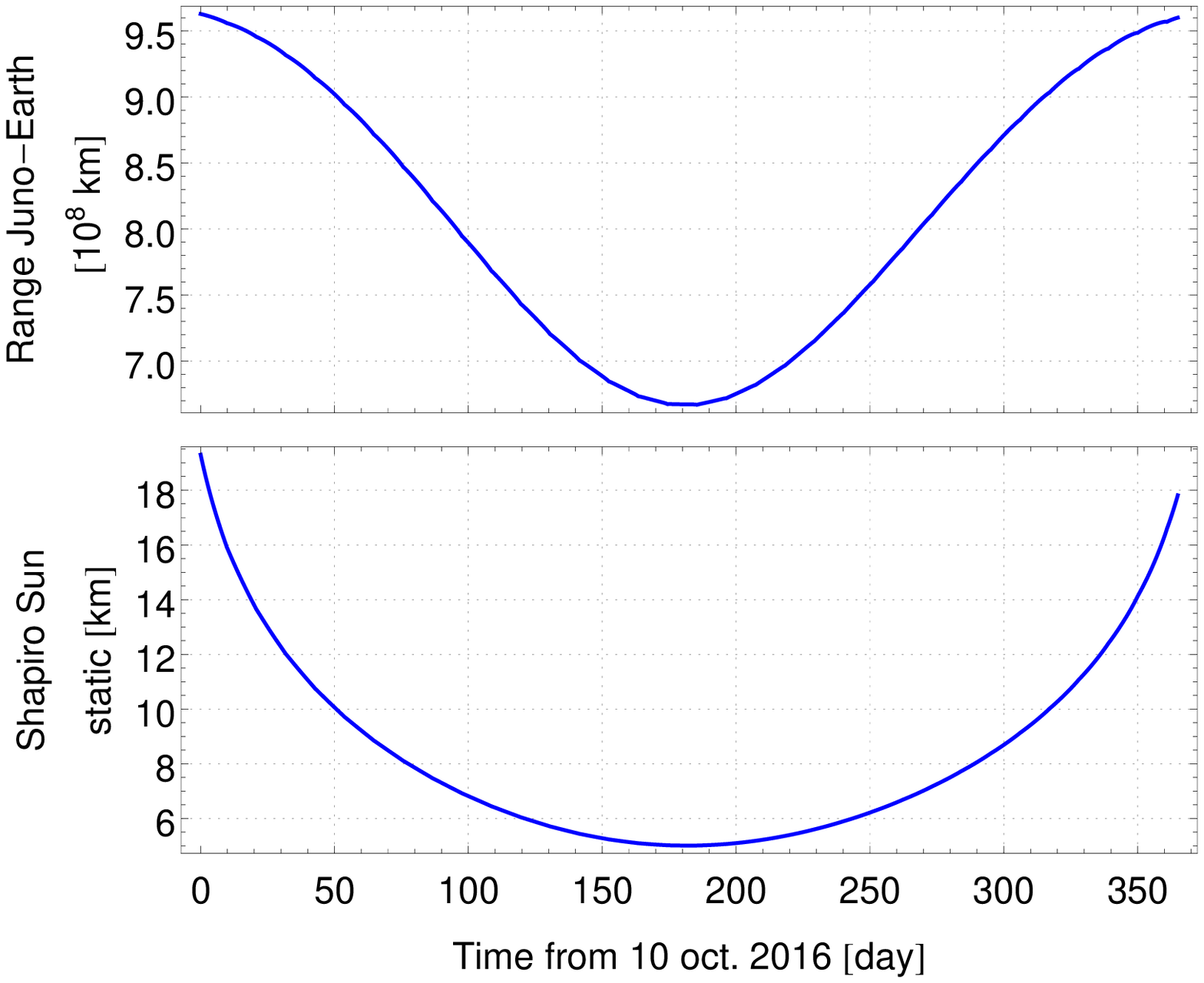}\hfill
\includegraphics[width=0.48\textwidth]{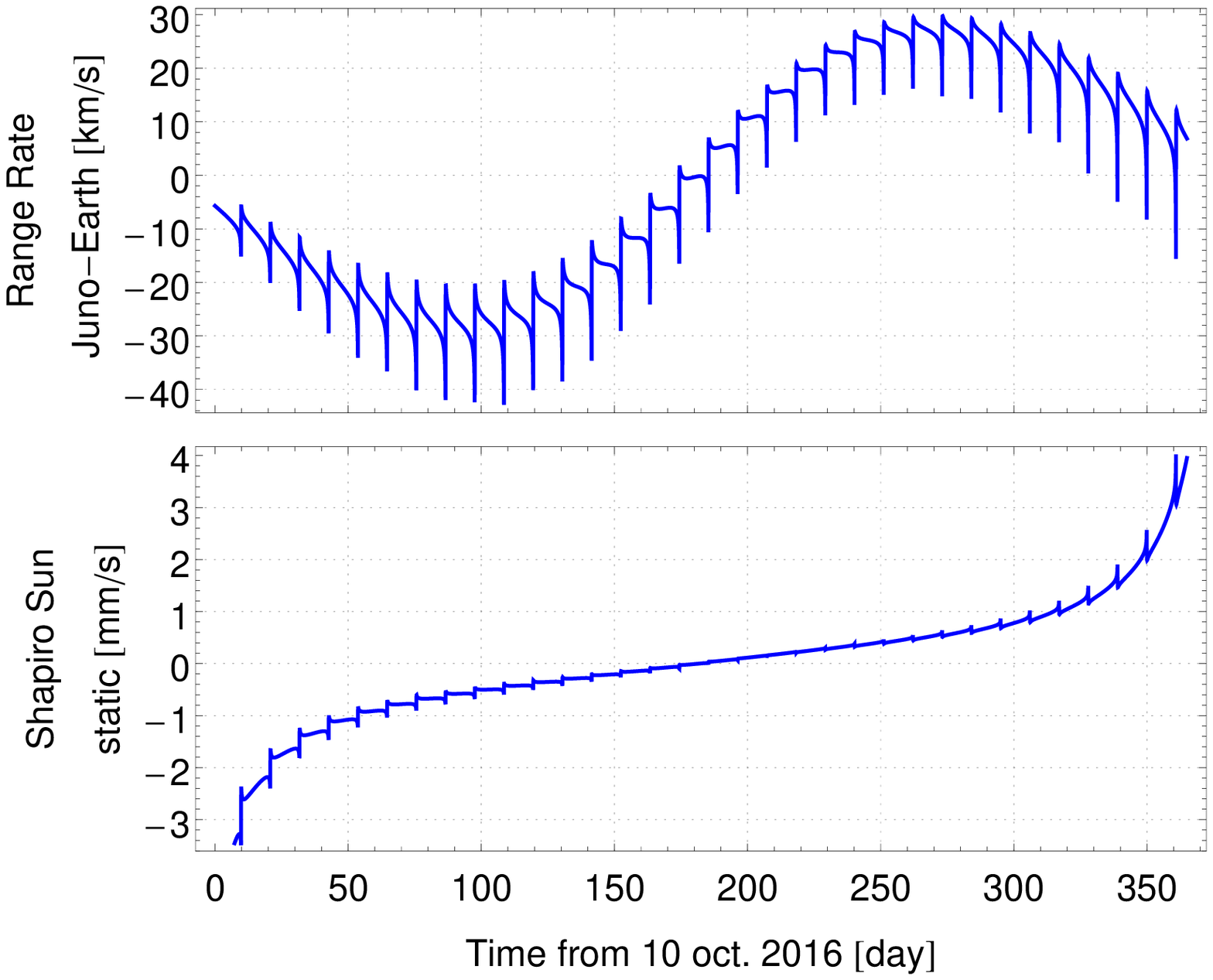}
\end{center}
\caption{Representation of different contributions on the range (left) and range-rate (right) between JUNO and Earth over one year. The contributions represented are the lower order contribution (the actual value of the observables) and the corrections produced by the Shapiro due to the monopole of the Sun.}
\label{fig:juno1}
\end{figure}

\begin{figure}[hbt]
\begin{center}
\includegraphics[width=0.48\textwidth]{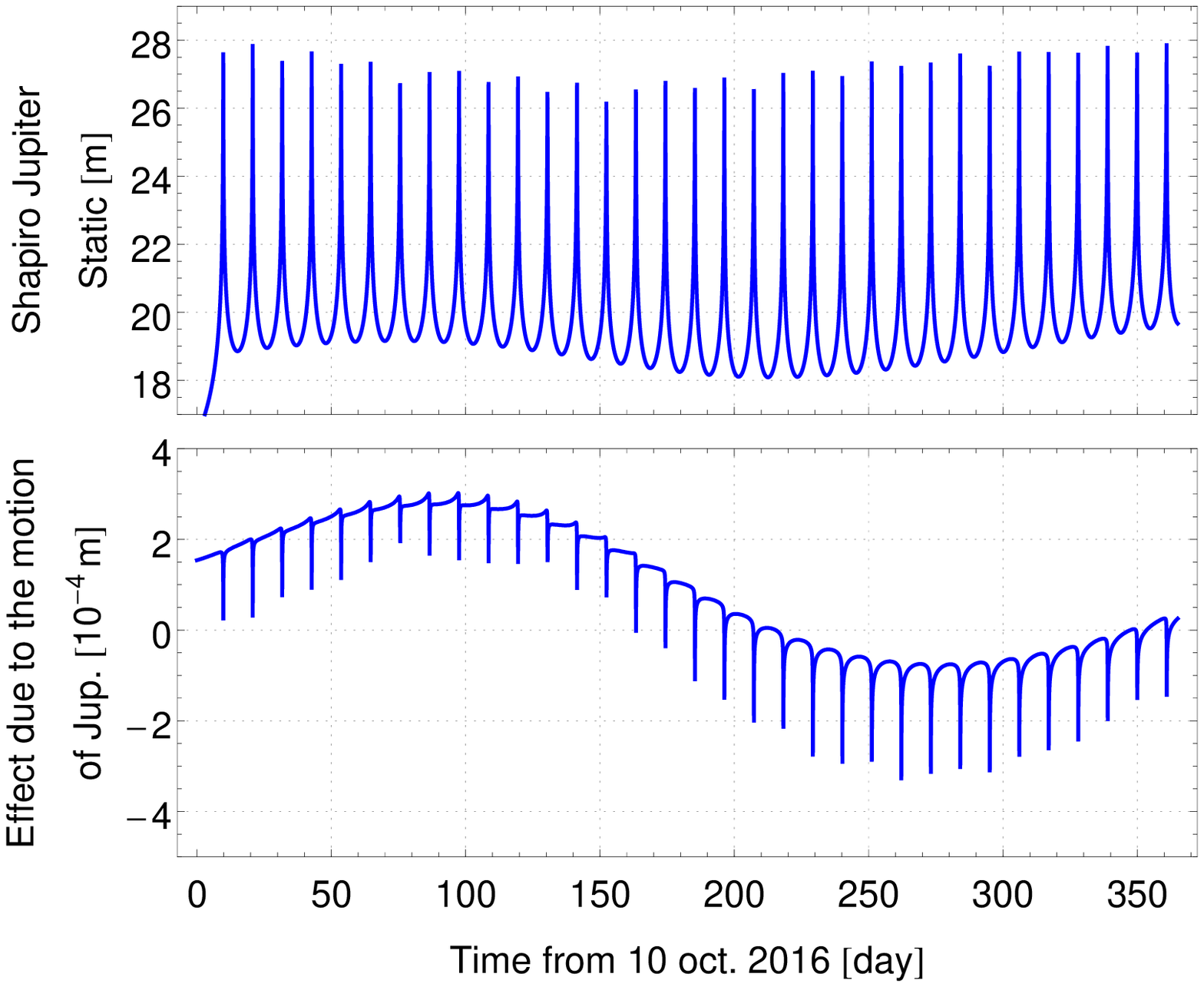}\hfill
\includegraphics[width=0.48\textwidth]{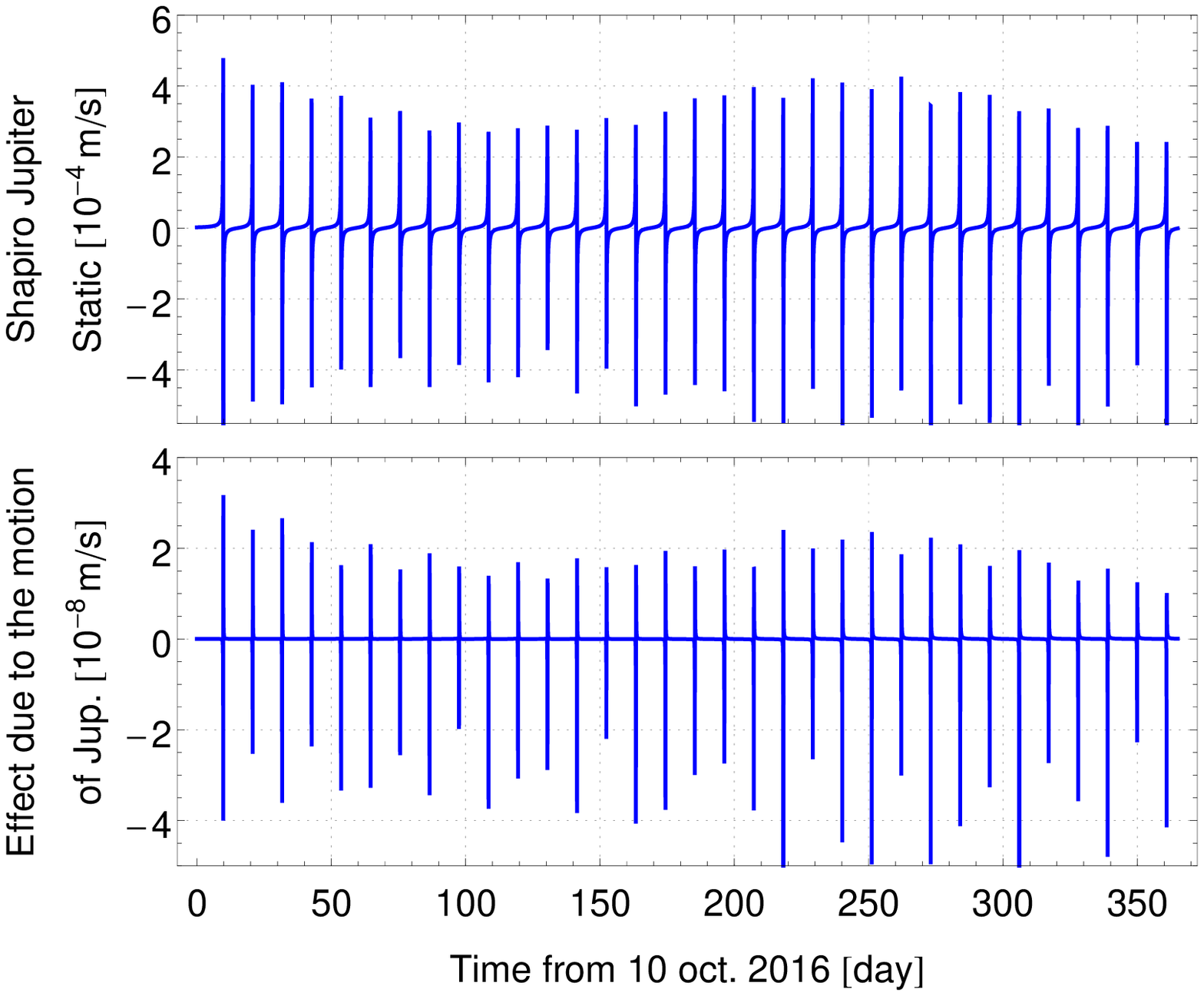}
\end{center}
\caption{Representation of different contributions on the range (left) and range-rate (right) between JUNO and Earth over one year. The contributions represented are the corrections produced by the Shapiro due to the monopole of the Jupiter (top) and the contributions due to the velocity of Jupiter (bottom).}
\label{fig:juno2}
\end{figure}

Fig.~\ref{fig:juno2} represents the contributions of the mass monopole of Jupiter on the range and on the range-rate. These contributions have been split into two parts: a part related to the case where Jupiter is static and a contribution proportional to Jupiter velocity $\beta_\textrm{Jup}$. The static part is computed using (\ref{eq:deltamono}) with the position of Jupiter taken at the critical time $t_0$ given by Eq.~\eqref{eq:t0}. The contribution relative to the velocity is computed by taking the difference between the relations (\ref{eq:deltamonomov}) and (\ref{eq:deltamono}). As one can see, the contributions relative to the motion of Jupiter are 2 orders of magnitude below JUNO expected accuracy and can safely be neglected in the modeling of the time transfer. A similar conclusion holds for the motion of the Sun around the Solar System barycenter which is even smaller. Note that the analytical results presented in these graphs have been checked by integrating numerically the TTF (\ref{eq:Deltanumeric}).

\begin{figure}[hbt]
\begin{center}
\includegraphics[width=0.48\textwidth]{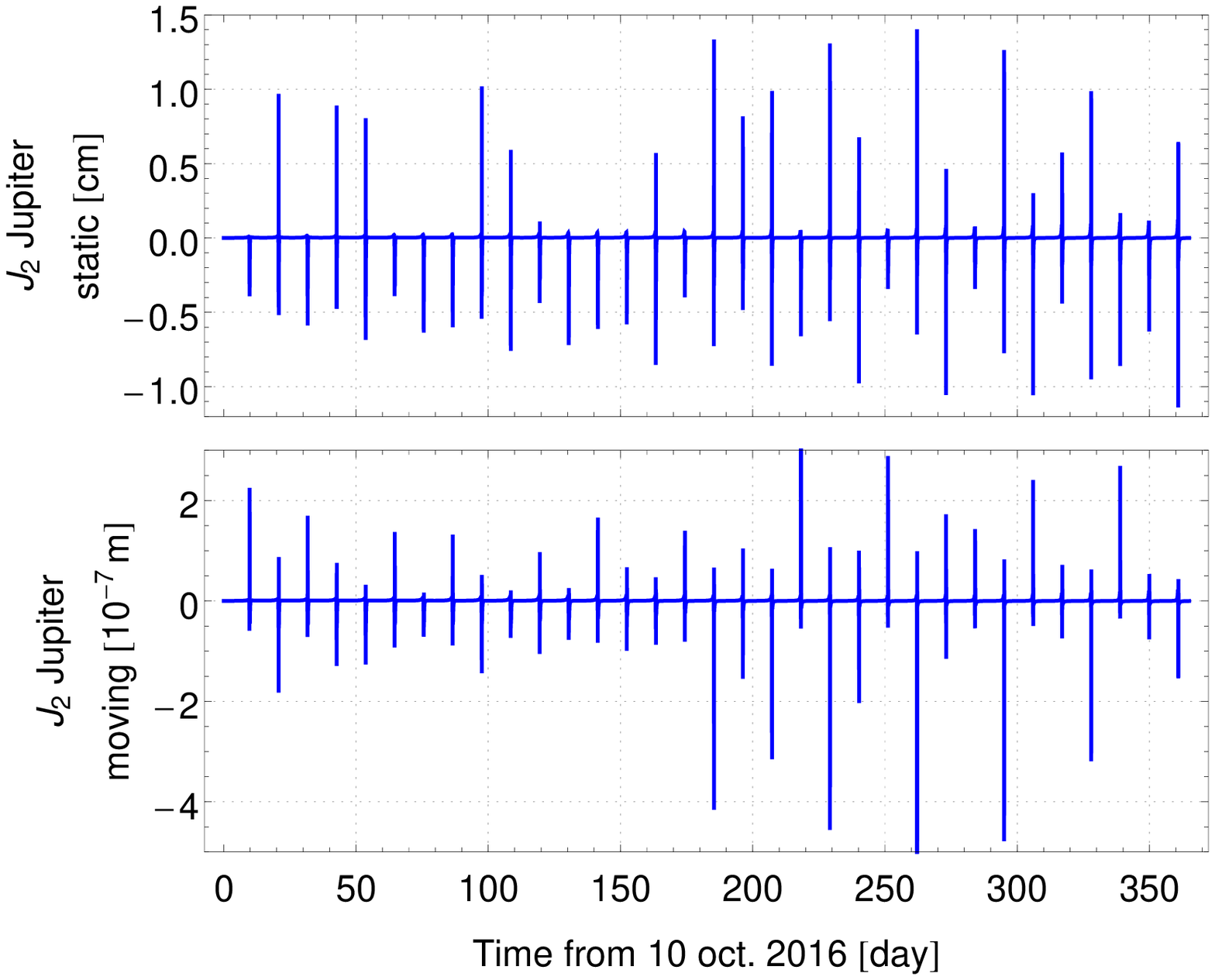}\hfill
\includegraphics[width=0.48\textwidth]{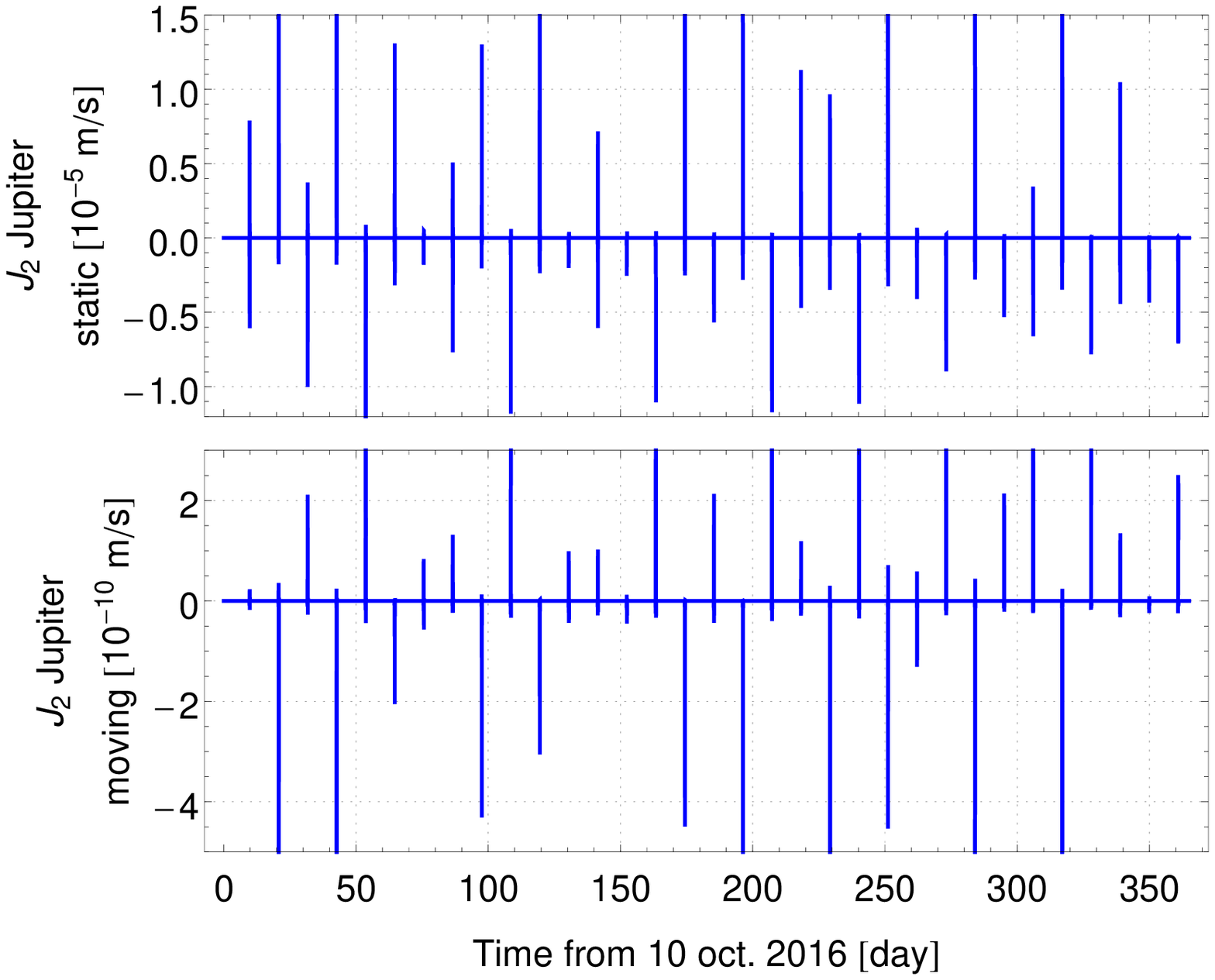}
\end{center}
\caption{Representation of different contributions on the range (left) and range-rate (right) between JUNO and Earth over one year. The contributions represented are the corrections produced by the $J_2$ of the Jupiter (top) and the contributions produced by the fact that the $J_2$ is moving.}
\label{fig:juno3}
\end{figure}

Fig.~\ref{fig:juno3} represents the contributions of the quadrupole of Jupiter ($J_2$) on the range and on the range-rate of JUNO. As above, we have split these contributions into two parts: one related to the case where Jupiter is static and one proportional to Jupiter velocity $\beta_\textrm{Jup}$. The static part is computed using (\ref{eq:J2stat}) with the position of Jupiter taken at the critical time $t_0$ given by Eq.~\eqref{eq:t0}. The contribution relative to the velocity is computed by taking the difference between the relations (\ref{eq:J2mov}) and (\ref{eq:J2stat}). As one can see, the contributions relative to the $J_2$ of Jupiter is of the same order as the expected JUNO's accuracy. Therefore the effect of the $J_2$ should be taken into account in the reduction of the tracking data. The contribution related to the velocity of the $J_2$ is far beyond the current tracking accuracy. Once again, the analytical results presented in these graphs have been checked by integrating numerically the TTF (\ref{eq:Deltanumeric}). It is important to notice that the curves depend highly on the geometry of the probe orbit. Since JUNO has a polar orbit and is never in conjunction with Jupiter, the velocity effects are not detectable. Therefore, the situation can be different for another space mission like JUICE~\cite{grasset:2013uq}.

\begin{figure}[hbt]
\begin{center}
\includegraphics[width=0.48\textwidth]{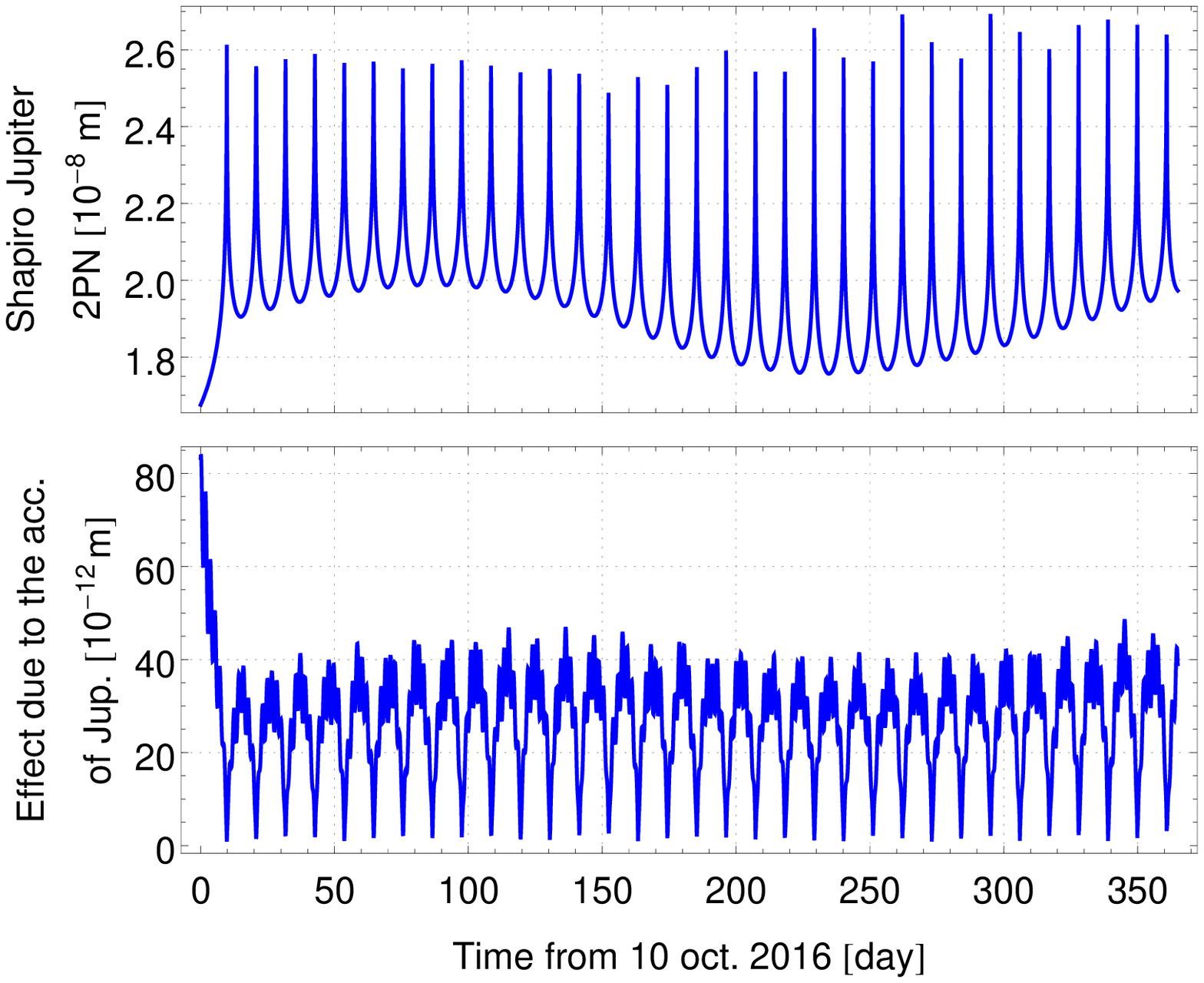}\hfill
\includegraphics[width=0.48\textwidth]{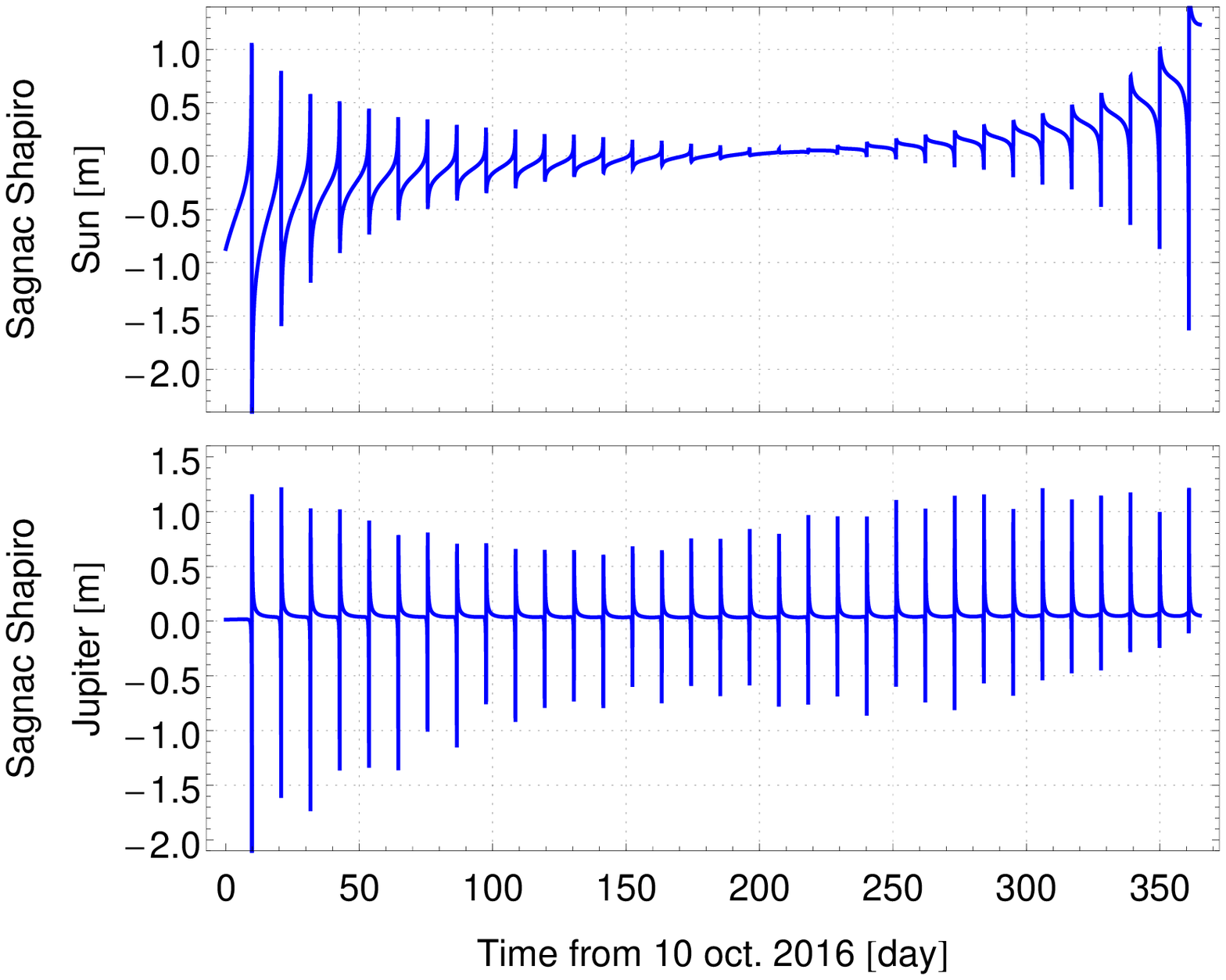}
\end{center}
\caption{Representation of different contributions on the range between JUNO and Earth over one year. Top Left: The 2PN contribution from the monopole of Jupiter (contribution proportional to $\beta_\textrm{Jup}^2$). Bottom Left: The contribution proportional to the acceleration of Jupiter. Right: the Sagnac contributions (proportional to the Sun mass and to Jupiter mass) due to the motion of JUNO.}
\label{fig:juno4}
\end{figure}

Fig.~\ref{fig:juno4} is given for illustrative purpose and shows more effects on the range of JUNO. First of all, the effects of the second order in $\beta_{\textrm{Jup}}$ are represented. It is computed by making the difference between the formula valid at all order in $\beta_{\textrm{Jup}}$ (\ref{eq:deltamonomov}) and the 1PN expansion (\ref{eq:deltap_pn}). This shows one can safely use the PN expansion presented in Section~\ref{sec:PNexp} within the Solar System. The effect of the acceleration of Jupiter on the range is also presented. This is computed by making the difference between the numerical integration of the TTF in which we are using the real Jupiter trajectory (\ref{eq:Deltanumeric}) and the result valid at all order in the velocity (\ref{eq:deltamonomov}). The small rapid oscillations come from oscillations in Jupiter acceleration, which results from the perturbations due to the Galilean satellites.

Finally, on the right of Fig.~\ref{fig:juno4} are represented the Sagnac effects due to the motion of JUNO. The contributions represented are due to the Shapiro of the Sun and Jupiter and it has been computed using (\ref{eq:sagnac}). These contributions should be included in the analysis of JUNO data either as a perturbation, either when solving the light-time iterations.

%%%%%%%%%%%%%%%%%%%%%%%%%%%%%%%%%%
\section{Conclusions}\label{sec:conclusion}
%%%%%%%%%%%%%%%%%%%%%%%%%%%%%%%%%%
In this paper, we compute the TTF and its derivatives in the field of uniformly moving axisymmetric bodies and in the field of arbitrarily moving point masses, which is useful in order to evaluate range, Doppler and astrometric observables. First, in Section~\ref{sec:genmetric} we compute a metric adapted to describe the space-time geometry due to $N$ bodies in a global reference system by using a Poincar\'e transformation. 

Then we present a general method to compute the TTF and its derivatives in the case where the bodies generating the gravitational field are in uniform motion. We show that the TTF in the case of uniform motion can be directly derived from the static TTF as can be seen from  Eq.~\eqref{eq:deltap3} and Eqs.~\eqref{eq:derdeltap2}. This result is very powerful and valid for any velocities. Moreover, in Section~\ref{sec:genMotion}, we have developed a general expression of the TTF in the case where the gravitational field is generated by arbitrarily moving point masses. The result is given as an integral over a straight line between the emitter and the receiver (\ref{eq:Deltanumeric}) which can be computed numerically. This general formulation has been used to numerically check our analytical derivations but is also useful to assess the effects due to the acceleration of the body on the light propagation.

Then, in Section~\ref{sec:movaxi} we show how our method can be easily applied to the metric presented in Section~\ref{sec:genmetric} to compute analytically the TTF and its derivatives (and thus the range, frequency-shift and astrometric direction) for a light signal propagating in the field of one or more axisymmetric bodies in uniform motion. The results of this paper complete the work of~\cite{le-poncin-lafitte:2008fk,bertone:2014uq} and in general extend the field of applicability of the TTF formalism~\cite{teyssandier:2008nx}.

Finally, as an example of our method, we compute the range and Doppler for the JUNO mission during its orbit around Jupiter and study in details the different perturbations due to the Sun and Jupiter on light propagation. In particular, we have shown that in addition to the standard Shapiro contributions due to the mass monopole of Jupiter and of the Sun, the contribution of Jupiter $J_2$ is also relevant at the level of accuracy expected for JUNO.  The motion of the Sun and of Jupiter produces effects too small compared to JUNO accuracy. Nevertheless, this conclusion depends highly on the geometry of JUNO orbit and it should be assessed carefully for other space mission (JUICE for example~\cite{grasset:2013uq}).

\begin{acknowledgments}
The authors thank the anonymous referee for useful remarks about the space-time metric used in this paper. The authors are grateful to the financial support of CNRS/GRAM and Observatoire de Paris/GPHYS.
\end{acknowledgments}

%%%%%%%%%%%%%%%%%%%%%%%
\bibliography{../../../JPL/JPL_byMe/biblio}
%%%%%%%%%%%%%%%%%%%%%%%%

\appendix

\section{Metric of arbitrarily moving point masses}\label{app:metricAcc}
The standard way to compute the metric for an arbitrarily moving point mass is to use the Li\'enard-Wiechert potentials as in \cite{kopeikin:1999kl,kopeikin:2002fr,zschocke:2014fk}. Based on the analogy between the Maxwell equations and the linearized Einstein field equations \cite{damour:1991tb}, it follows the guidelines of classical electromagnetism (see Chapter 8 of \cite{landau:1975fk}). This procedure is described in details in \cite{zschocke:2014fk}. According to the formulas for the retarded potentials, the field at the point of observation at time $t$ is determined by the state of motion of the body at the earlier time $t_r$ which is determined by (\ref{eq:retarded}) (in the following, the index $r$ denotes quantities evaluated at the retarded time $t_r$).

We can introduce a reference system comoving with the body at the retarded time and whose temporal origin coincide with the retarded time. The coordinates with respect to this frame will be denoted by $X^\alpha$ and they can be derived by the instantaneous Lorentz transformation 
\begin{equation}\label{eq:local_transfo}
	X^\alpha = \tilde \Lambda ^\alpha_{r\ \mu}(x^\mu-x^\mu_{pr})
\end{equation}
where $x^0_{pr}=ct_r$ and $\bx_{pr}=\bx_p(t_r)$ are the coordinates of the body at the retarded time. It is important to notice that the Lorentz transformation is done at the retarded time (i.e. $\tilde \Lambda ^\alpha_{r\ \mu}=\tilde \Lambda ^\alpha_{\mu}(t_r)$). The 4-vector $X^\alpha$ is a null 4-vector ($\eta_{\alpha\beta}X^\alpha X^\beta=\eta_{\mu\nu}(x^\mu-x^\mu_r)(x^\nu-x^\nu_r)=c^2(t-t_r)^2-|\bx -\bx_p(t_r)|=0$). In this frame, the space-time metric is known $H^{\alpha\beta}=\frac{2W(X^i)}{c^2}\delta^{\alpha\beta}$ which can be written in a manifestly covariant way as $H^{\alpha\beta}=\frac{4W(X^i)}{c^2}(U_{pr}^\alpha U_{pr}^\beta-\frac{1}{2}\eta^{\alpha\beta})$ with $U_{pr}^\alpha$ the 4-velocity of the body at the retarded time (in the comoving frame $U_{pr}^\alpha=\delta^{\alpha 0}$). Since the expression of the metric is manifestly covariant, we can express it in the global frame by using the local transformation (\ref{eq:local_transfo})
\begin{equation}
	h^{\mu\nu}=\frac{4W(X^i)}{c^2}\left(u_{pr}^\alpha u_{pr}^\beta - \frac{1}{2}\eta^{\alpha\beta} \right)
\end{equation}
where the $X^i$ are given by the transformation (\ref{eq:local_transfo}) and where $u^{\mu}_{rp}$ is now given by $u^{\mu}_{rp}=\gamma_{pr} (1 ,\bm \beta_{pr})$ with $\gamma_{pr}=\gamma_p(t_r)$ and $\bv_{pr}=\left.d\bx_p/dt\right|_{t_r}$. 

We can write the space-time metric as (\ref{eq:metricacc}) with $X^i$ given by (\ref{eq:local_transfo}) which can be explicitly written as (\ref{eq:transfAcc}).
For example, in the case of a point mass ($W=GM/R$), one gets
$$
R=|X^i|=\gamma_{pr}\left(r_{pr}-(\br_{pr}\cdot\bm \beta_{pr})\right)
$$
and the metric can be written as
\begin{subequations}\label{eq:metricAccP}
	\begin{eqnarray}
		h^{00}&=&\frac{2GM}{c^2\left(r_{pr}-(\br_{pr}\cdot\bm \beta_{pr})\right)}\gamma_{pr}(1+\beta_{pr}^2) + \mathcal O(G^2)\; ,\\
		h^{0i}&=&\frac{4GM}{c^2\left(r_{pr}-(\br_{pr}\cdot\bm \beta_{pr})\right)}\beta_{pr}^i\gamma_{pr} + \mathcal O(G^2) \; , \\
		h^{ij}&=&\frac{2GM}{c^2\left(r_{pr}-(\br_{pr}\cdot\bm \beta_{pr})\right)\gamma_{pr}}(\delta_{ij}+2\beta_{pr}^i\beta_{pr}^j\gamma_{pr}^2) + \mathcal O(G^2) .
	\end{eqnarray}
\end{subequations}
This expression is exactly the same as the one found in \cite{kopeikin:1999kl,kopeikin:2002fr,zschocke:2014fk}. In the limit of small velocities, the expressions of the IAU metric is recovered (see Appendix~\ref{app:limitiau2}).

\section{Correspondence with the IAU metric}
\subsection{Case of uniformly moving bodies}\label{app:limitiau1}
In Sec.~\ref{sec:metricUnif}, we derive the post-Minkowskian metric related to uniformly moving bodies (\ref{eq:metric1}). It is interesting to show that the post-Minkowskian limit of the IAU metric \cite{soffel:2003bd} is recovered in the limit of the small velocities. In order to show this, we first need to develop the argument appearing in the potential $W$ from Eq.~\eqref{eq:XLW} as
\begin{eqnarray}
	\tilde \Lambda^i_{\mu}(x^\mu-b^\mu)=-\gamma_p \beta^i_p c (t-t_0)+x^i-x^i_p(t_0)+\frac{\gamma_p^2}{1+\gamma_p}\beta_p^i \bm \beta \cdot (\bx - \bx_p(t_0)).
\end{eqnarray}
Using the fact that the motion of the body is uniform, this expression can also be written as
\begin{equation}
	\tilde \Lambda^i_{\mu}(x^\mu-b^\mu)=x^i-x^i_p(t)+\frac{\gamma_p^2}{1+\gamma_p}\beta_p^i \bm \beta \cdot (\bx - \bx_p(t)).
\end{equation}
In the limit of the small velocities ($\beta_p<<1$), we have
\begin{equation}
	\tilde \Lambda^i_{\mu}(x^\mu-b^\mu)=x^i-x^i_p(t)+\frac{1}{2}\beta_p^i \bm \beta_p \cdot (\bx - \bx_p(t))+\mathcal O(\beta_p^4).
\end{equation}
Using this expansion in the limit of small velocities, the expression of the potential $W$ appearing in the metric (\ref{eq:metric1}) becomes 
\begin{eqnarray}
	W\left[\tilde \Lambda^i_{\mu}(x^\mu-b^\mu)\right]=W\left[x^i-x^i_p(t)\right]+\frac{1}{2}W_{,j}\left[x^i-x^i_p(t)\right]\beta^j \bm \beta_p \cdot (\bx - \bx_p(t))+\mathcal O(\beta_p^4) \; ,
\end{eqnarray}
where $W_{,j}=\partial W/\partial X^j$. Introducing this expression in the metric (\ref{eq:metric1}) leads to
\begin{subequations}
	\begin{eqnarray}
		h^{00}&=&\frac{2W(x^i-x^i_p)}{c^2}+4\frac{W(x^i-x^i_p)}{c^2}\beta_p^2 + \frac{\bm \beta_p \cdot (\bx - \bx_p(t))}{c^2} \beta_p^j W_{,j}(x^j-x^j_p(t)) + \mathcal O(G^2) +\mathcal O(\beta_p^4/c^2)\; ,\\
		h^{0i}&=&\frac{4W(x^i-x^i_p)}{c^2}\beta_p^i + \mathcal O(G^2) +\mathcal O(\beta_p^3/c^2)\; , \\
		h^{ij}&=&\frac{2W(x^i-x^i_p)}{c^2}\delta_{ij} + \mathcal O(G^2) +\mathcal O(\beta_p^2/c^2).
	\end{eqnarray}
\end{subequations}
For example, in the case of a point mass ($W=\frac{GM}{R}$), the last expression becomes 
\begin{subequations}\label{eq:iaumetricU}
	\begin{eqnarray}
		h^{00}&=&2\frac{GM}{r_p c^2}+4\frac{GM}{r_pc^2}\beta_p^2 -\frac{GM}{c^2r_p^3} (\br\cdot\bm \beta_p)^2+ \mathcal O(G^2) +\mathcal O(\beta_p^4/c^2)\; ,\\
		h^{0i}&=&4\frac{GM}{r_p c^2}\beta_p^i + \mathcal O(G^2) +\mathcal O(\beta_p^3/c^2)\; , \\
		h^{ij}&=&2\frac{GM}{r_p c^2}\delta_{ij} + \mathcal O(G^2) +\mathcal O(\beta_p^2/c^2).
	\end{eqnarray}
\end{subequations}
with $r_p=|\br_p|$ and $\br_p=\bx-\bx_p(t)$. This expression is exactly the one recommended in the IAU conventions (see Eqs.~(8),  and (51-55) from \cite{soffel:2003bd} or the Resolutions B1.5. in the Appendix of the same paper).

\subsection{Case of arbitrarily moving point masses}\label{app:limitiau2}
The metric~(\ref{eq:metricacc}) or (\ref{eq:metricAccP}) describes the space-time geometry around an arbitrarily moving point mass at the first post-Minkowskian approximation.  It is interesting to show that the  post-Minkowskian limit of the IAU metric \cite{soffel:2003bd} is recovered in the limit of the small velocities.

We need to express the quantities at the retarded time $t_r$ as a function of the quantities at the time $t$. Since 
\begin{equation}
	t_r-t=-\frac{r_{pr}}{c},
\end{equation}
we have
\begin{eqnarray}
	\br_p (t)&=&\br_p=\bx-\bx_p(t)=\bx-\bx_p(t_r)-(t-t_r)\bv_p(t_r)-\frac{(t-t_r)^2}{2}\ba_p(t_r)+\mathcal O\left((t-t_r)^3\right)\\
		&=& \br_{pr}-\frac{r_{pr}}{c}\bv_{pr}-\frac{r_{pr}^2}{2c^2}\ba_{pr}+\mathcal O(c^{-3}).
\end{eqnarray}
It is useful to notice that in the last term of this expression, we can replace $t_r$ by $t$ (this will introduce a higher order correction). A simple calculation leads to
\begin{equation}
	r_p^2+(\bm \beta_p\cdot \br_p)^2=\left(r_{pr}-(\bm \beta_{pr}\cdot \br_{pr})\right)^2+r_{pr}^2\beta_{pr}^2- \frac{r_p^2}{c^2}\ba_p \cdot \br_p +\mathcal O(c^{-3}).
\end{equation}
This leads to
\begin{equation}
	r_{pr}-(\bm \beta_{pr}\cdot \br_{pr})=r_p\left[1-\frac{\beta_{pr}^2}{2}+\frac{(\bm \beta_{p}\cdot\br_{p})^2}{2r_p^2}+\frac{\ba_p \cdot \br_p}{2c^2}\right]+\mathcal O(c^{-3}).
\end{equation}
Since 
\begin{equation}
	\gamma_{pr}=\gamma_p+\mathcal O(c^{-3}) = 1+\frac{\beta_p^2}{2}+\mathcal O(c^{-3}),
\end{equation}
we have
\begin{equation} \label{eq:factgamma}
	\frac{1}{\gamma_{pr}\left(r_{pr}-(\bm \beta_{pr}\cdot \br_{pr})\right)}=\frac{1}{r_p}\left[1-\frac{(\bm\beta_p\cdot\br_p)^2}{2r_p^2}-\frac{\ba_p\cdot\br_p}{2c^2}\right].
\end{equation}
Introducing Eq.~\eqref{eq:factgamma} in the space-time metric~(\ref{eq:metricAccP}) leads to
\begin{subequations}
	\begin{eqnarray}
		h^{00}&=&\frac{2GM}{c^2r_p} + 4\frac{GM}{c^2r_p}\beta_p^2 -\frac{GM}{c^2r_p^3} (\br_p\cdot\bm \beta_p)^2 - \frac{GM}{c^4 r_p} (\ba_p\cdot\br_p)+ \mathcal O(G^2) + \mathcal (1/c^5)\; ,\\
		h^{0i}&=&4\frac{GM}{r_p^2c^2}\beta_p^i + \mathcal O(G^2) +\mathcal O(1/c^4)\; , \\
		h^{ij}&=&2\frac{GM}{r_pc^2}\delta_{ij} + \mathcal O(G^2) +\mathcal O(1/c^3).
	\end{eqnarray}
\end{subequations}
The only additional term with respect to the metric (\ref{eq:iaumetricU}) is the term proportional to the acceleration in $g_{00}$. This term is exactly the one appearing in the IAU metric (see Eq.~(54) of \cite{soffel:2003bd}) as already noticed in \cite{eddington:1938jk}.

\end{document}